\documentclass[12pt,final]{article}

 \usepackage[T1]{fontenc}
   \usepackage{fourier}
\usepackage{xcolor}
\usepackage[reqno]{amsmath}
\usepackage{nicefrac}
\usepackage{tikz}
\usetikzlibrary{arrows}
\usetikzlibrary{trees,calc,snakes,fit,shapes,positioning,automata,quotes}
\usetikzlibrary{decorations.pathreplacing}
\usetikzlibrary{decorations.pathmorphing}
\usetikzlibrary{decorations.markings}
\usetikzlibrary{decorations.shapes}
\tikzset{
  treenode/.style = {align=center, inner sep=0pt, text centered,
    font=\sffamily},
  arn_n/.style = {treenode, circle, black, font=\sffamily\bfseries, draw=black,
    fill=white, text width=1.5em},
  arn_r/.style = {treenode, circle, bg, draw=bg, 
    text width=1.5em},
  arn_x/.style = {treenode, circle, purple,font=\sffamily\bfseries,
   text width=2em}
}
\tikzset{
  invisible/.style={opacity=0},
  visible on/.style={alt={#1{}{invisible}}},
  alt/.code args={<#1>#2#3}{%
    \alt<#1>{\pgfkeysalso{#2}}{\pgfkeysalso{#3}} 
  },
}
\tikzset{
    photon/.style={decorate, decoration={snake}, draw=red}}
\tikzset{electron/.style={draw=blue, thick, postaction={decorate},decoration={markings,mark=at position .75 with {\arrow[draw=blue]{>}}}}}
\tikzset{electron2/.style={draw=red,very thick, postaction={decorate},decoration={markings,mark=at position .55 with {\arrow[draw=red,very thick]{>}}}}}
\tikzset{gluon/.style={->,thick,decorate, draw= mLightBrown,
        decoration={coil,amplitude=4pt, segment length=5pt}}}
 \tikzset{gluon2/.style={thick,decorate, draw=mLightBrown,
        decoration={coil,amplitude=4pt, segment length=5pt}}}
\tikzset{nero/.style={decorate,draw=black}}
\tikzset{bianco/.style={decorate,draw=bg}}
 \tikzset{ every node/.style={inner sep=0pt,minimum size=1mm},
  nsnode/.style={draw,circle,black,minimum size=2mm},
  asnode/.style={draw,circle,myblue,fill=myblue},
  bsnode/.style={draw,circle,black,fill=black,nero},
  csnode/.style={draw,circle,red,fill=red, minimum size=2mm},
  every fit/.style={inner sep=-1.5pt,text width=1cm}  }
\usepackage{soul}
\usepackage{bbm}
\usepackage{amsthm}
\makeatletter
\def\thm@space@setup{%
  \thm@preskip=\parskip \thm@postskip=0pt
}
\makeatother
\makeatletter
\newcommand{\leqnomode}{\tagsleft@true}
\newcommand{\reqnomode}{\tagsleft@false}
\makeatother
\usepackage{setspace}
\usepackage{amsfonts}
\usepackage{graphicx}
\usepackage{subfig}
\usepackage{fnpct}
\usepackage{apptools}
\usepackage[justification=centering]{caption}
\usepackage{natbib}
\usepackage{titlesec}

\usepackage[colorlinks,linkcolor=links,citecolor=cites,urlcolor=MyDarkBlue]{hyperref}
\usepackage{cleveref}
\usepackage{xr}
\usepackage{geometry}
\usepackage{enumitem}
\usepackage[section]{placeins}
\definecolor{MyDarkBlue}{rgb}{0,0.08,0.45}
\definecolor{cites}{HTML}{324b13}
\definecolor{links}{HTML}{1a663b}
\definecolor{MyLightBlue}{cmyk}{0.1,0.8,0,0.1}
\newtheoremstyle{ex}
{1pt}
{1pt}
{}
{}
{\bfseries}
{.}
{.5em}
{}%
\theoremstyle{ex}\newtheorem{ex}{Example}
\theoremstyle{theorem}\newtheorem{theorem}{Theorem}
\theoremstyle{theorem}\newtheorem{definition}{Definition}
\theoremstyle{theorem}\newtheorem{prop}{Proposition}
\theoremstyle{theorem}\newtheorem{rem}{Remark}
\theoremstyle{theorem}
\theoremstyle{theorem}

\newcommand{\posteriorsale}{\ensuremath{F_{2S}}}
\newcommand{\posteriordelay}{\ensuremath{F_{2D}}}

\newcommand{\cut}{\ensuremath{\hat{\type}}}

\newcommand{\newcut}{\ensuremath{\tilde{\type}}}

\newcommand{\maxt}{\ensuremath{\overline{\type}}}
\newcommand{\mint}{\ensuremath{\underline{\type}}}

\newcommand{\dateindexb}{\ensuremath{\tau}}
\newcommand{\terminal}{\ensuremath{T}}

\newcommand{\type}{\ensuremath{\theta}}

\newcommand{\typec}{\ensuremath{\tilde{\type}}}
\newcommand{\typed}{\ensuremath{\type^\prime}}

\newcommand{\Types}{\ensuremath{\Theta}}

\newcommand{\hag}{\ensuremath{h_A}}
\newcommand{\Hag}{\ensuremath{H_A}}
\newcommand{\hagt}{\ensuremath{\hag^t}}
\newcommand{\Hagt}{\ensuremath{\Hag^t}}

\newcommand{\Hagterminal}{\ensuremath{\Hag^{\terminal+1}}}

\newcommand{\hagtc}{\ensuremath{\tilde{h}_A^t}}

\newcommand{\Public}{\ensuremath{H}}
\newcommand{\Publict}{\ensuremath{\Public^t}}

\newcommand{\publict}{\ensuremath{h^t}}
\newcommand{\publictminus}{\ensuremath{h^{t-1}}}

\newcommand{\publictplus}{\ensuremath{h^{t+1}}}

\newcommand{\publicdateindexb}{\ensuremath{h^\dateindexb}}

\newcommand{\prior}{\ensuremath{\mu_1}}
\newcommand{\posterior}{\ensuremath{\mu}}
\newcommand{\posteriorb}{\ensuremath{\mu^\prime}}
\newcommand{\beliefend}{\ensuremath{\mu}}
\newcommand{\beliefendb}{\ensuremath{\beliefend^\prime}}

\newcommand{\beliefbeg}{\ensuremath{\nu}}

\newcommand{\mechanism}{\ensuremath{\mathbf{M}}}
\newcommand{\mechanismt}{\ensuremath{\mechanism_t}}
\newcommand{\mechanismtminus}{\ensuremath{\mechanism_{t-1}}}

\newcommand{\mechanismb}{\ensuremath{\mechanism^\prime}}

\newcommand{\mechanismbt}{\ensuremath{\mechanismb_t}}

\newcommand{\betat}{\ensuremath{\beta^{\mechanismt}}}

\newcommand{\betabt}{\ensuremath{\beta^{\mechanismbt}}}

\newcommand{\varphim}{\ensuremath{\varphi^\mechanism}}
\newcommand{\varphit}{\ensuremath{\varphi^{\mechanismt}}}

\newcommand{\inputm}{\ensuremath{M^\mechanism}}
\newcommand{\inputt}{\ensuremath{M^{\mechanismt}}}

\newcommand{\outputm}{\ensuremath{S^\mechanism}}
\newcommand{\outputt}{\ensuremath{S^{\mechanismt}}}

\newcommand{\alphat}{\ensuremath{\alpha^{\mechanismt}}}

\newcommand{\alphabt}{\ensuremath{\alpha^{\mechanismbt}}}

\newcommand{\zpt}{\ensuremath{z_{(s_t,a_t)}(\mechanismt)}}

\newcommand{\tripletfp}{\ensuremath{\type,\hagt,\mechanismt}}

\newcommand{\plainstrat}{\ensuremath{\sigma}}

\newcommand{\stratp}{\ensuremath{\sigma_P}}
\newcommand{\stratpt}{\ensuremath{\sigma_{Pt}}}

\newcommand{\stratpb}{\ensuremath{\stratp^\prime}}
\newcommand{\strat}{\ensuremath{\sigma_A}}
\newcommand{\stratt}{\ensuremath{\sigma_{At}}}

\newcommand{\stratb}{\ensuremath{\strat^\prime}}

\newcommand{\comm}{\ensuremath{r}}
\newcommand{\commb}{\ensuremath{\comm^\prime}}
\newcommand{\participate}{\ensuremath{\pi}}
\newcommand{\participateb}{\ensuremath{\participate^\prime}}

\newcommand{\assessment}{\ensuremath{(\stratp,\strat,\beliefend)}}
\newcommand{\assessmentb}{\ensuremath{(\stratpb,\stratb,\beliefendb)}}

\newcommand{\action}{\ensuremath{a}}
\newcommand{\actiont}{\ensuremath{\action_t}}
\newcommand{\actiontminus}{\ensuremath{\action_{t-1}}}

\newcommand{\outsideoption}{\ensuremath{\action^*}}

\newcommand{\actionuptot}{\ensuremath{\action^{t-1}}}

\newcommand{\allocations}{\ensuremath{A}}
\newcommand{\feasibleallocations}{\ensuremath{\pazocal{A}}}
\newcommand{\Posteriors}{\ensuremath{\Delta(\Types)}}

\newcommand{\measurablet}{\ensuremath{\widetilde{\Types}}}

\newcommand{\measurablea}{\ensuremath{\widetilde{A}}}
\newcommand{\measurablem}{\ensuremath{\widetilde{U}}}

\newcommand{\polish}{\ensuremath{Y}}
\newcommand{\polishb}{\ensuremath{X}}

\newcommand{\genericb}{\ensuremath{x}}

\newcommand{\polishm}{\ensuremath{\Delta(\polish)}}

\newcommand{\cor}{\ensuremath{\omega}}
\newcommand{\corb}{\ensuremath{\cor^\prime}}
\newcommand{\Cor}{\ensuremath{\Omega}}

\newcommand{\kernel}{\ensuremath{\kappa}}

\newcommand{\Mechanisms}{\ensuremath{\pazocal{M}}}

\newcommand{\canonical}{\ensuremath{\Mechanisms_C}}
\newcommand{\signal}{\ensuremath{s}}

\newcommand{\publicttau}{\ensuremath{\publicdateindexb_t}}
\newcommand{\publicttplus}{\ensuremath{\publictplus_t}}
\newcommand{\Privatettau}{\ensuremath{H_{At}^\dateindexb}}

\newcommand{\priorf}{\ensuremath{F_1}}
\newcommand{\priorpdf}{\ensuremath{f_1}}
\newcommand{\posteriorf}{\ensuremath{F_2}}
\newcommand{\posteriorfb}{\ensuremath{\posteriorf^\prime}}

\newcommand{\transfer}{\ensuremath{x}}

\usepackage{calrsfs}
\DeclareMathAlphabet{\pazocal}{OMS}{zplm}{m}{n}
\usepackage{upgreek}
\AtAppendix{\counterwithin{lemma}{section}}
\AtAppendix{\counterwithin{prop}{section}}
\AtAppendix{\counterwithin{corollary}{section}}
\AtAppendix{\counterwithin{definition}{section}}
\AtAppendix{\counterwithin{equation}{section}}
\AtAppendix{\counterwithin{rem}{section}}
\newcommand{\Belief}{\ensuremath{B}}

\newcommand{\virtual}{\ensuremath{J}}

\newcommand{\objective}{\ensuremath{H}}
\newcommand{\priceopt}{\ensuremath{p_{2\delta}^*}}

\newcommand{\cutopt}{\ensuremath{\cut_{\delta}^*}}

\newcommand{\eqbmoutcomes}{\ensuremath{\pazocal{O}^*}}
\newcommand{\ceqbmoutcomes}{\ensuremath{\pazocal{O}^C}}

\newcommand{\signals}{\ensuremath{S}}
\newcommand{\messages}{\ensuremath{M}}

\newcommand{\spot}{\ensuremath{\varphi}}
\newcommand{\spotb}{\ensuremath{\spot^\prime}}

\newcommand{\spotbt}{\ensuremath{\spot_t^\prime}}

\newcommand{\cspot}{\ensuremath{\spot^C}}

\newcommand{\cspotdateindexb}{\ensuremath{\cspot_\dateindexb}}
\newcommand{\cspottt}{\ensuremath{\left(\cspotdateindexb\right)_{\dateindexb\geq t}}}

\newcommand{\spotindexb}{\ensuremath{\spot_\dateindexb}}
\newcommand{\spotbindexb}{\ensuremath{\spotb_\dateindexb}}
\newcommand{\spott}{\ensuremath{(\spotindexb)_{\dateindexb\geq t}}}

\newcommand{\spotbtt}{\ensuremath{(\spotbindexb)_{\dateindexb\geq t}}}

\newcommand{\spotbttplus}{\ensuremath{(\spotbindexb)_{\dateindexb\geq t+1}}}

\newcommand{\extensive}{\ensuremath{\Gamma}}

\newcommand{\signalt}{\ensuremath{\signal_t}}

\newcommand{\agentstrat}{\ensuremath{\strat}}
\newcommand{\agentbelief}{\ensuremath{\mu}}
\newcommand{\agentstratb}{\ensuremath{\agentstrat^\prime}}
\newcommand{\agentbeliefb}{\ensuremath{\agentbelief^\prime}}

\newcommand{\agentassessment}{\ensuremath{(\strat,\posterior)}}
\newcommand{\agentassessmentb}{\ensuremath{(\stratb,\posteriorb)}}

\newcommand{\doutcome}{\ensuremath{\eta}}
\newcommand{\doutcomeb}{\ensuremath{\doutcome^\prime}}
\newcommand{\Deviant}{\ensuremath{\mathsf{D}}}

\newcommand{\priort}{\ensuremath{\posterior_t}}
\newcommand{\allocationsuptot}{\ensuremath{\allocations^{t-1}}}

\newcommand{\nmssg}{\ensuremath{\emptyset}}
\newcommand{\nsignal}{\ensuremath{\emptyset}}

\newcommand{\saempty}{\ensuremath{\signals_j\allocations_\emptyset}}
\newcommand{\saemptyb}{\ensuremath{\signals\allocations_\emptyset}}
\newcommand{\msaempty}{\ensuremath{\messages_i\signals_j\allocations_\emptyset}}

\newcommand{\collection}{\ensuremath{\pazocal{I}}}

\newcommand{\allocationsterminal}{\ensuremath{\allocations^{\terminal}}}

\newcommand{\deviation}{\ensuremath{d}}

\newcommand{\gamecollection}{\ensuremath{G_\collection}}
\newcommand{\mechanismscollection}{\ensuremath{\Mechanisms_\collection}}
\newcommand{\periodtwo}{\ensuremath{t=2}}
\newcommand{\periodone}{\ensuremath{t=1}}
\newcommand{\Mechanismsb}{\ensuremath{\Mechanisms^\prime}}

\newcommand{\periodgeqone}{\ensuremath{t\geq 1}}

\makeatletter
\renewcommand\section{\@startsection {section}{1}{\z@}%
                                   {-1.5ex \@plus -1ex \@minus -.2ex}%
                                   {1.5ex \@plus.2ex}%
                                   {\normalfont\scshape}}
\renewcommand\subsection{\@startsection{subsection}{2}{\z@}%
                                     {-1.5ex\@plus -1ex \@minus -.2ex}%
                                     {0.25ex \@plus .2ex}%
                                     {\normalfont\itshape}}
\makeatother

\title{  Mechanism Design with Limited Commitment\thanks{We thank the Editor and four anonymous referees for excellent comments. We would also like to thank Rahul Deb, Fran\c{c}oise Forges, David Miller, Dan Quigley, Luciano Pomatto, Pablo Schenone, Omer Tamuz, and especially Michael Greinecker and Max Stinchcombe, as well as various audiences for thought-provoking questions and illuminating discussions.  Alkis Georgiadis-Harris, Nathan Hancart, and Ignacio N\'u\~nez provided excellent research assistance. Vasiliki Skreta is grateful for generous financial support through the ERC consolidator grant 682417 ``Frontiers in design.'' This research is supported by grants from the National Science Foundation (SES-1851744 and SES-1851729).}}
\author{Laura Doval\thanks{Columbia University and CEPR. E-mail: \href{mailto:laura.doval@columbia.edu}{\texttt{laura.doval@columbia.edu}}}\and Vasiliki Skreta\thanks{University of Texas at Austin, University College London, and CEPR. E-mail: \href{mailto:vskreta@gmail.com}{\texttt{vskreta@gmail.com}}}}

\begin{document}
\pagenumbering{gobble}
\maketitle
\begin{abstract}
We develop a tool akin to the revelation principle for dynamic mechanism-selection games in which the designer can only commit to short-term mechanisms. We identify a \emph{canonical} class of mechanisms rich enough to replicate the outcomes of any equilibrium in a mechanism-selection game between an uninformed designer and a privately informed agent. A cornerstone of our methodology is the idea that a mechanism should encode not only the rules that determine the allocation, but also the information the designer obtains from the interaction with the agent. Therefore, how much the designer learns, which is the key tension in design with limited commitment, becomes an explicit part of the design. Our result simplifies the search for the designer-optimal outcome by reducing the agent's behavior to a series of participation, truthtelling, and Bayes' plausibility constraints the mechanisms must satisfy.

\end{abstract}
\small
\textsc{Keywords:} \emph{mechanism design, limited commitment, revelation principle, information design, short-term mechanisms}\\
\textsc{JEL classification:} D84, D86
\normalsize
\newpage
\clearpage
\pagenumbering{arabic}
\section{Introduction}
The standard assumption in dynamic mechanism design is that the designer can commit to long-term contracts. This assumption is useful: it allows us to characterize the best possible payoff for the designer in the presence of adverse selection and/or moral hazard, and it is applicable in many settings. Often, however, this assumption is made for technical convenience. Indeed, when the designer can commit to long-term contracts, the mechanism-selection problem can be reduced to a constrained optimization problem thanks to the \emph{revelation principle}.\footnote{The ``revelation principle'' denotes a class of results in mechanism design; see, for instance, \cite{myerson1982optimal}.}
However, as the literature starting with \cite{freixas1985planning} and \cite{laffont1988dynamics} shows, when the designer can commit only to short-term contracts, the tractability afforded by the revelation principle is lost. Indeed, mechanism design problems with limited commitment are difficult to analyze without imposing auxiliary assumptions either on the class of contracts available to the designer, as in  \cite{acharya2017progressive} and \cite{gerardi2020dynamic}, or on the length of the horizon, as in \cite{skreta2006sequentially,skreta2015optimal}.

This paper provides a ``revelation principle'' for dynamic mechanism-selection games in which the designer can only commit to short-term mechanisms. We study a class of \emph{mechanism-selection} games between an uninformed designer and an informed agent with persistent private information. Although the designer can commit within each period to the terms of the interaction--the current mechanism--he cannot commit to the terms the agent faces later on, namely, the mechanisms that are chosen in the continuation game. First, we identify a set of mechanisms, and hence \emph{a} mechanism-selection game, that is sufficient to replicate any equilibrium outcome of any mechanism-selection game. Second, we identify a set of strategies for the designer and the agent that is sufficient to replicate any equilibrium outcome of the identified mechanism-selection game. We illustrate how our result can be used to characterize the designer's optimal outcome as the solution to a constrained optimization problem that only involves the designer. In this problem, the designer chooses amongst mechanisms that satisfy the usual truthtelling and participation constraints, and a third constraint, which captures the designer's sequential rationality.

The starting point of our analysis is the class of mechanisms we allow the designer to choose from. Following \cite{myerson1982optimal} and \cite{bester2007contracting}, we consider mechanisms as illustrated in Figure \ref{fig:general}:

\begin{figure}[h!]
\begin{minipage}{0.5\textwidth}
\begin{center}
\subfloat[General Mechanisms \textcolor{white}{line break}]{
\begin{tikzpicture}[scale=1,thick]
\node[label=left:{\Types\ }](agent) at (0,0){};
\node[label=right:{$M$}](input) at (2,0){};
\node[label=right:{$S\times \allocations$}](output) at (4,0){};
\node (place) at (2.5,0){};
\draw[->](agent)edge node[above]{{\small$r(\cdot|\type)$}}(input);
\draw[->](place)edge node[above]{{\small$\varphi(\cdot|m)$}}(output);
\node[label=left:{}](agent1) at (0,-0.25){};
\node[label=right:{}](input1) at (2,-0.25){};
\node[label=right:{}](output1) at (4,-0.25){};
\node (place1) at (2.5,-0.25){};
\end{tikzpicture}\label{fig:general}}
\end{center}
\end{minipage}
\begin{minipage}{0.5\textwidth}
\begin{center}
\subfloat[Canonical Mechanisms: $M=\Types,S=\Posteriors$]{
\begin{tikzpicture}[scale=1,thick]
\node[label=left:{\Types\ }](agent) at (0,0){};
\node[label=right:{$M$}](input) at (2,0){};
\node[label=right:{$S$}](output) at (4,0){};
\node[label=right:{\allocations\ }](allocations) at (6,0){};
\node (place) at (2.5,0){};
\node (place2) at (4.5,0){};
\draw[->](agent)edge node[above]{{\small$r(\cdot|\type)$}}(input);
\draw[->](place)edge node[above]{{\small$\beta(\cdot|m)$}}(output);
\draw[->](place2) edge node[above]{{\small$\alpha(\cdot|s)$}}(allocations);
\node[label=left:{}](agent1) at (0,-0.25){};
\node[label=right:{}](input1) at (2,-0.25){};
\node[label=right:{}](output1) at (4,-0.25){};
\node (place1) at (2.5,-0.25){};
\end{tikzpicture}\label{fig:canonical}}
\end{center}
\end{minipage}
\caption{Mechanisms}\label{fig:mechanism}
\end{figure}
\newpage
Having observed her private information (her type, $\type\in \Types$), the agent privately reports an \emph{input} message, $m\in M$, to the mechanism, which then determines the distribution, $\varphi(\cdot|m)$, from which an \emph{output} message, $s\in S$, and an allocation, $a\in\allocations$, are drawn. 
The output message and the allocation are \emph{publicly observable}: they constitute the contractible parts of the mechanism.

When the designer has commitment, the standard revelation principle implies that, without loss of generality, we can restrict attention to mechanisms satisfying three properties: (i) $M=\Types$, (ii) $|\messages|=|\signals|$, and (iii) $\varphi$ is such that by observing the output message, the designer learns the input message, in this case, the agent's type report. Moreover, the revelation principle implies we can restrict attention to equilibria in which the agent truthfully reports her type, which means the designer not only learns the agent's type report upon observing the output message but also learns the agent's true type.

Why restricting attention to mechanisms that satisfy properties (i)-(iii) and to truthtelling equilibria is with loss of generality under limited commitment is therefore clear: upon observing the output message, the designer learns the agent's type report and hence her type. Then, the agent may have an incentive to misreport if the designer cannot commit to not react to this information. \label{page-BS}This intuition is behind the main result in \cite{bester2001contracting}, which is the first paper to provide a general analysis of optimal mechanism design with limited commitment. The authors restrict attention to mechanisms such that the cardinality of the set of input and output messages is the same, and $\spot$ is such that, by observing the output message, the designer learns the input message.\footnote{\label{footnote-mech-class}The class of mechanisms considered in \cite{bester2001contracting} encompasses the mechanisms considered by most papers in the literature on limited commitment starting from \cite{laffont1988dynamics}.} 
They show that to sustain payoffs in the Pareto frontier, mechanisms in which input messages are type reports are without loss of generality. However, focusing on truthtelling equilibria is with loss of generality. In a follow-up paper, \cite{bester2007contracting} lift the restrictions on the class of mechanisms (i.e., (ii) and (iii) above) and show in a one-period model that focusing on mechanisms in which input messages are type reports and on truthtelling equilibria is without loss of generality. The authors, however, do not characterize the set of output messages.
Whether taking the set of input messages to be the set of type reports is without loss when the designer and the agent interact repeatedly is also unclear (see the discussion after \autoref{theorem:main-theorem}).

The main contribution of this paper is to show that, under limited commitment, it is without loss of generality to take the set of output messages to be the set of the designer's posterior beliefs about the agent's type; that is, $S=\Posteriors$. \autoref{theorem:main-theorem} identifies a set of mechanisms, and hence a mechanism-selection game, that is enough to replicate any equilibrium outcome of any mechanism-selection game in which the designer chooses mechanisms as in \autoref{fig:general}. In this game, which we denote the \emph{canonical game}, the designer can only offer mechanisms in which input messages are type reports and output messages are beliefs. Moreover, \autoref{theorem:main-theorem} shows that any equilibrium of the canonical game can be replicated by a \emph{canonical equilibrium} in which the agent always participates in the mechanisms offered in equilibrium by the designer, and input and output messages have a \emph{literal} meaning: the agent truthfully reports her type, and if the mechanism outputs a given posterior, this posterior coincides with the designer's equilibrium belief  about the agent's type. Furthermore, in a canonical equilibrium, the designer only offers the agent \emph{canonical mechanisms}, in which conditional on the output message, the allocation is  drawn independently of the agent's type report (see \autoref{fig:canonical}). Thus, like the standard revelation principle, \autoref{theorem:main-theorem} implies that to characterize the distributions over types and allocations that can be achieved in some equilibrium in some mechanism-selection game, it is without loss of generality to restrict attention to the analysis of the canonical equilibria of the canonical game.

 \label{page-th-simpl}\autoref{theorem:main-theorem}  provides researchers with a tractable way to analyze problems of mechanism design with limited commitment by making how much the designer learns about the agent an explicit part of the design. A major challenge in the received literature on limited commitment is how to keep track of how the agent's best response to the mechanism affects the information that the designer obtains from the interaction, which in turn affects the designer's incentives to offer the mechanism in the first place. Instead, our framework allows us to reduce the agent's best response to the designer's mechanism and its informational feedback to a familiar set of constraints that the mechanism must satisfy: the participation and incentive compatibility constraints for the agent, and the Bayes' plausibility constraint. This avoids having to consider complicated mixed strategies on the part of the agent (see \citealp{laffont1988dynamics,bester2001contracting}) and transforms it instead into a program that combines elements of mechanism design and information design. Indeed, we exploit the information design elements to derive properties of the mechanisms (see \autoref{prop:finite-support} and our companion work, \citealp{doval2020optimal,doval2021purchase}).

We prove \autoref{theorem:main-theorem} under the assumption that the set of types is at most countable and extend \autoref{theorem:main-theorem} to the case in which the agent's type is drawn from a continuum (a leading case in mechanism design) in  \autoref{theorem:main-theorem-continuum}. As we explain in \autoref{sec:complications}, the results in \cite{aumann1961borel,aumann1964mixed} imply that with a continuum type space we cannot rely on the usual formulation of an extensive-form game. This is the reason that we first conduct our analysis under the assumption that the set of types is at most countable, deriving \autoref{theorem:main-theorem} in the standard game-theoretic framework. \autoref{sec:dynamic-mechanisms} then develops a new framework that circumvents the issues raised by Aumann, while allowing us to extend  \autoref{theorem:main-theorem} to continuum type spaces. The framework is based on the idea that any fixed sequence of mechanisms determines a well-defined extensive-form game for the agent. Like in the mechanism-selection game, \autoref{theorem:main-theorem-continuum} shows it is without loss of generality to assume the designer offers the agent sequences of canonical mechanisms and to restrict attention to canonical equilibria  within the extensive-form game defined by the mechanisms.

We apply our results to a seemingly well-understood problem and show that our tools can shed new light on it.\footnote{\autoref{theorem:main-theorem} also opens the door to the analysis of optimal mechanisms under limited commitment in infinite-horizon settings. See \cite{doval2020optimal}, where we solve an infinite-horizon binary-type version of the sale of a durable good.}\label{page-ration-intro} As in \cite{skreta2006sequentially}, \autoref{ex:sdg} considers a seller, who owns one unit of a durable good, and interacts over two periods with a buyer with persistent and private information. In \autoref{ex:sdg} the seller offers canonical mechanisms, whereas in  \cite{skreta2006sequentially} the seller offers mechanisms in the class considered by \cite{laffont1988dynamics} and \cite{bester2001contracting}.
 In \autoref{sec:sdg-program}, we show how \autoref{theorem:main-theorem} can be used to reduce the characterization of the seller's maximum revenue to the characterization of the solution to a constrained optimization problem (see \ref{eq:seller-opt}). Furthermore, in \autoref{sec:sdg}, we argue that the solution to the program \ref{eq:seller-opt} also describes the seller's maximum revenue when the buyer's type is drawn from a continuum. We then
%
show how to obtain the envelope representation of the agent's payoffs and hence, the dynamic virtual surplus representation of the seller's payoff. As we explain in \autoref{sec:sdg}, characterizing the solution to \ref{eq:seller-opt} is outside the scope of this paper.\label{page-not-solve-intro}  Instead, we use the virtual surplus representation of the seller's payoff to show that, in contrast to the main result in \cite{skreta2006sequentially}, the seller can do strictly better than in the optimal posted-price mechanism.\footnote{\autoref{rem:vs} discusses how canonical mechanisms differ from the mechanisms in \cite{skreta2006sequentially}, which explains the difference in the results.} Starting from the optimal posted-price mechanism, we show that the seller has a deviation to an alternative mechanism that combines posted prices with a form of \emph{rationing}. This allows us to connect the mechanism design literature on the sale of a durable good with the work in theoretical industrial organization on alternative strategies for a durable goods monopolist, such as rationing (\citealp{denicolo1999rationing}).

By highlighting the canonical role of beliefs as the signals employed by the mechanism, \autoref{theorem:main-theorem} underscores the importance of jointly determining the mechanism together with how information is used in the mechanism and transmitted across periods. In doing so, it marries information design, which studies the design of information structures in a given institution, with mechanism design, which generally studies institutional design within a given information structure. 
\textbf{Related Literature:} The paper contributes to the literature on mechanism design with limited commitment with an informed agent with persistent private information, referenced throughout the introduction.\footnote{A designer's lack of commitment can take various forms that are not considered in this paper but have been studied in others. See, for instance, \cite{mcadams2007credible}, \cite{vartiainen2013auction}, and \cite{akbarpour2020credible}, in which the designer cannot commit even to obeying the rules of the \emph{current} mechanism.} Following the seminal contribution of \cite{bester2001contracting}, a body of work studies optimal mechanisms under limited commitment in finite-horizon settings with finitely many types (e.g., \citealp{bisin2006markets,hiriart2011weak,fiocco2015consumer,beccuti2018dynamic}).  
However, the results in \cite{bester2001contracting} do not extend to settings with a continuum of types and/or infinite horizon. On the one hand, the proof strategy in \cite{bester2001contracting} relies on the assumption of finitely many types. On the other hand, their result only applies
 if the designer is earning his highest payoff consistent with the agent's payoff (see Lemma 1 in \citealp{bester2001contracting}). Thus, implicit in their multistage extension is a restriction to equilibria of the mechanism-selection game that possess a Markov structure, which, as shown by \cite{ausubel1989reputation}, may not be enough to characterize the designer's best equilibrium payoff in infinite-horizon settings. As a consequence, there is a small body of work that studies mechanism-selection games with a continuum of types and finite horizon (\citealp{skreta2006sequentially,skreta2015optimal,deb2015dynamic}), or in infinite-horizon settings, imposing restrictions on the class of contracts that can be offered (e.g., \citealp{acharya2017progressive,gerardi2020dynamic}), or on the solution concept (e.g., \citealp{acharya2017progressive}). 

Due to the difficulties with the revelation principle, a large body of work in public finance, political economy, and taxation considers optimal time-consistent policies in settings where private information is fully non-persistent (see \citealp{sleet2008politically, farhi2012non, golosov2021social}). Moreover, a large literature studies the effect of limited commitment within a specific class of mechanisms: price dynamics in the durable goods literature (\citealp{bulow1982durable,gul1986foundations,stokey1981rational})  and reserve price dynamics in auction settings  (\citealp{mcafee1997sequentially, liu2019auctions}).



By highlighting the role that the designer's beliefs about the agent play in mechanism design with limited commitment, our paper also relates to \cite{lipnowski2020cheap} and \cite{best2017persuasion}, who study models of direct communication between an informed sender and an uninformed receiver.

\textbf{Organization:} The rest of the paper is organized as follows. \autoref{sec:model} describes the model and notation and \autoref{sec:main} introduces the main theorem for at most countable type spaces. \autoref{sec:continuum} extends our analysis to continuum type spaces. Throughout the paper, we use a two-period version of the model in \cite{skreta2006sequentially} to illustrate our results. \autoref{sec:conclusions} concludes and discusses further directions. Omitted statements and all proofs are in the appendix (Appendices \ref{appendix:definitions}-\ref{appendix:sdg}) and the online appendix, \cite{supplement}. 

\section{Model}\label{sec:model}

\textbf{Primitives:} Two players, a principal (he) and an agent (she), interact over $\terminal\leq\infty$ periods.\footnote{To simultaneously analyze the cases of finite and infinite horizon, we abuse notation as follows. When $\terminal=\infty$, and notation of the form $\periodone,\dots, \terminal$, $\sum_{\periodone}^\terminal$, or $\times_{\periodone}^\terminal$, appears, we take this to mean $\periodone, \dots$, $\sum_{t\in\mathbb{N}}$, or $\times_{t\in\mathbb{N}}$, respectively. \label{ftn:terminal-notation}} Before the game starts, the agent observes
her type, $\type\in\Types$, which is distributed according to a full-support distribution \prior. We initially assume \Types\ is at most countable and consider continuum type spaces in \autoref{sec:continuum}. Each period, as a result of the interaction between the principal and the agent, an allocation $\action\in\allocations$ is determined. Let $\allocationsterminal$ denote the set $\times_{\periodone}^\terminal\allocations$. When the agent's type is \type\ and the allocation is $\action^{\terminal}\in\allocationsterminal$, the principal and the agent's payoffs are given by $W(\action^{\terminal},\type)$ and $U(\action^{\terminal},\type)$, respectively.

We allow for the possibility that past allocations influence what the principal can offer the agent in the future. Thus, for each \periodgeqone, a correspondence $\feasibleallocations_t:\allocations^{t-1}\rightrightarrows\allocations$ exists
such that for every sequence of allocations up to period $t$, $\actionuptot=(\action_1,\dots,\action_{t-1})$, $\feasibleallocations_t(\actionuptot)$ describes the set of allocations the principal can offer in period $t$ (with the convention that when $t=1$, $\action^0=\{\emptyset\}$). Furthermore, we assume an allocation \outsideoption\ exists such that \outsideoption\ is always available. Below, allocation \outsideoption\ plays the role of the agent's outside option. Given the general structure of payoffs, it is without loss of generality to assume that \outsideoption\ is time-independent.

We impose some technical restrictions on our model.\footnote{In what follows, we adopt the following notational conventions. First, all Polish spaces are endowed with their Borel $\sigma$-algebra. Second, product spaces are endowed with their product $\sigma$-algebra. Third, for a Polish space, \polish, we let \polishm\ denote the set of all Borel probability measures over \polish, endowed with the weak$^*$ topology. Thus,  \polishm\ is also a Polish space (\citealp{aliprantis2013infinite}). For any two measurable spaces, \polishb\ and \polish, a transition probability from \polishb\ to \polish\ is a measurable function $\zeta:\polishb\mapsto\Delta(\polish)$. When integrating under the measure $\zeta(\genericb)$, we use the notation $\zeta(\cdot|\genericb)$. \label{ftn:mathematical-conventions}} The sets \Types\ and \allocations\ are Polish, that is, completely metrizable, separable, topological spaces. They are endowed with their Borel $\sigma$-algebra. We also assume \Types\ is compact. Endowing product sets with their product $\sigma$-algebra, we assume the principal and the agent's utility functions, $W$ and $U$, are bounded measurable functions. Similarly, for each \periodgeqone\ and for each $\actionuptot\in\allocations^{t-1}$, the set $\feasibleallocations_t(\actionuptot)$ is a measurable set. \label{page-meas-corr} 
 
 \textbf{Mechanisms:} In each period, the allocation is determined by a mechanism $\mechanismt=(\inputt,\outputt,\varphit)$, where \inputt\ and \outputt\ are the mechanism's input and output messages and $\varphit$ assigns to each $m\in\inputt$ a distribution over $\outputt\times\allocations$. 
%
 We endow the principal with a collection  $\{(M_i,S_i)\}_{i\in\collection}$ of input and output message sets, such that (i) $\messages_i,\signals_i$ are Polish spaces, (ii) $|\Types|\leq|M_i|$, $M_i$ is at most countable, and (iii) $|\Posteriors|\leq|S_i|$. Moreover, we assume $(\Types,\Posteriors)$ is an element in that collection. Let $\mechanismscollection$ denote the set of all mechanisms with message sets $(M_i,S_i)_{i\in\collection}$ that is,
$\mechanismscollection=\cup_{i,j\in\collection}\left\{\spot:\messages_i\mapsto\Delta(\signals_j\times\allocations):\spot\text{ is measurable }\right\}$. 

Three remarks are in order. First, we restrict the principal to choosing mechanisms in \mechanismscollection. This restriction allows us to have a well-defined strategy space for the principal, thereby avoiding set-theoretic issues related to self-referential sets.\footnote{To be precise, the set of all mechanisms is not constructible under Zermelo-Fraenkel's set theory, with the axiom of choice, since its axioms preclude Russell's paradox.\label{ftn:zfc}} The analysis that follows shows the choice of the collection plays no further role in the analysis. Second, because each $\messages_i$ is at most countable, the set of mechanisms $\mechanismscollection$ is a Polish space.\label{page-mechanisms-polish} As we discuss in \autoref{sec:continuum}, this property is key to being able to define a mechanism-selection game for a given collection \collection\ (see also \autoref{ftn:mechanisms-histories}). Third, we note all aspects of the environment, except the agent's type $\type\in\Types$, are common knowledge between the principal and the agent.

%

%
%
\textbf{Mechanism-selection game(s):} \label{page-mech-sel-game}Each collection \collection\ induces a mechanism-selection game, which we denote by \gamecollection, and is defined as follows. At the beginning of each period, both players observe the realization of a public randomization device, $\omega\sim U[0,1]$. The principal then offers the agent a mechanism, \mechanismt, with the property that for all $m\in\inputt$, $\varphit(\outputt\times\feasibleallocations_t(\actionuptot)|m)=1$, where recall that $\actionuptot$ describes the allocations implemented through period $t-1$. Observing the mechanism, the agent decides whether to participate in the mechanism ($\participate=1$) or not ($\participate=0$). If she does not participate in the mechanism, \outsideoption\ is implemented and the game proceeds to period $t+1$. Instead, if she chooses to participate, she sends a message $m\in\inputt$, which is unobserved by the principal. An output message and an allocation $(\signalt,\actiont)$ are drawn according to $\varphit(\cdot|m)$. The output message and the allocation are observed by both the principal and the agent, and the game proceeds to period $t+1$.

\textbf{Histories:} The game \gamecollection\ has two types of histories: public and private. Public histories capture what the \emph{principal} knows through period $t$: the past realizations of the public randomization device, his past choices of mechanisms, the agent's participation decisions, and the realized output messages and allocations. We let \publict\ denote a public history through period $t$ and let \Publict\ denote the set of all such histories. Instead, private histories capture what the \emph{agent} knows through period $t$. First, the agent knows the public history of the game and her input messages into the mechanism (henceforth, the agent history). Second, the agent also knows her private information. We let \hagt\ denote an agent's history through period $t$ and let $\Hagt(\publict)$ denote the set of agent histories consistent with public history \publict. Thus, $\Types\times\Hagt(\publict)$ denotes the set of private histories consistent with public history \publict.

\textbf{Belief system and strategies:} Private histories capture what the principal does not know about the agent in period $t$: he is uncertain about both the agent's payoff-relevant type, \type, and the agent history, \hagt. Thus, a belief for the principal in period $t$ at public history \publict\ is a distribution $\beliefend_t(\publict)\in\Delta(\Types\times\Hagt(\publict))$. The collection $(\beliefend_t)_{\periodone}^T$ denotes the belief system.

A behavioral strategy\label{page-strategies} for the principal is a collection of measurable mappings $(\stratpt)_{\periodone}^T$, where for each period $t$ and each public history \publict, \stratpt(\publict) describes the principal's (possibly random) choice of mechanism at \publict.\footnote{Because the set \mechanismscollection\ is Polish, the public and private histories are Polish, because they are the (at most) countable product of Polish spaces.\label{ftn:mechanisms-histories}} Similarly, a behavioral strategy for the agent is a collection of measurable mappings $(\stratt)_{\periodone}^T\equiv(\participate_t,\comm_t)_{\periodone}^T$, where for each period $t$, each private history $(\type,\hagt)$, and each mechanism, \mechanismt, $\participate_t(\type,\hagt,\mechanismt)$ describes the agent's participation decision, whereas $\comm_t(\type,\hagt,\mechanismt)$ describes the agent's choice of input messages in the mechanism, conditional on participation. 

The tuple $(\stratp,\strat,\beliefend)\equiv(\stratpt,\stratt,\beliefend_t)_{\periodone}^T$ defines an \emph{assessment}. 

\textbf{Equilibrium:} For each collection \collection, we study the equilibria of the mechanism-selection game $\gamecollection$. By equilibrium, we mean \emph{Perfect Bayesian equilibrium} (henceforth, PBE), informally defined as follows. An assessment \assessment\ is a PBE if it is sequentially rational and the belief system satisfies Bayes' rule where possible.  
The formal statement is in \autoref{appendix:definitions}. For now, we note that if the principal's strategy space is finite, \Types\ is finite, and the mechanisms used by the principal have finite support, our definition of PBE coincides with that in \cite{fudenberg1991perfect}.\footnote{The only difference between Bayes' rule where possible  (\autoref{definition:brwp}) and consistency in sequential equilibrium is that under PBE, the principal can assign zero probability to a type and then, after the agent deviates, can assign positive probability to that same type.\label{ftn:pbe}}

\textbf{Equilibrium outcomes:} The prior \prior\ together with a strategy profile $(\stratp,\strat)$ determine a distribution over the terminal nodes $\Types\times\Hagterminal$. We are interested instead in the distribution they induce over the payoff-relevant outcomes, $\Types\times\allocationsterminal$.  We say $\doutcome\in\Delta(\Types\times\allocationsterminal)$ is a PBE outcome if a PBE of the mechanism-selection game exists that induces \doutcome. We denote by $\eqbmoutcomes_\collection$ the set of PBE outcomes of $\gamecollection$.

Throughout, we use the following example to illustrate the concepts in the paper:
\begin{ex}\label{ex:sdg}
A seller (the principal) and a buyer (the agent) interact over two periods; that is, $\terminal=2$. \label{page-ex} The seller owns one unit of a durable good and assigns value $0$ to it. The buyer's value for the good, denoted by $\type\in\Types$, is her private information. We denote by \prior\ the seller's prior belief over \Types\ and by \maxt\ the maximum element of \space\Types. An allocation is a pair $(q,\transfer)\in\{0,1\}\times\mathbb{R}\equiv\allocations$, where $q$ indicates whether the good is sold ($q=1$) or not ($q=0$), and \transfer\ is a payment from the buyer to the seller. If the good is sold in period $1$, the game ends; that is, $\feasibleallocations_2\left((1,\transfer)\right)=\{0\}\times\mathbb{R}$ and $\feasibleallocations_2\left((0,\transfer)\right)=\allocations$. Moreover, if the buyer rejects the mechanism, the good is not sold and no payments are made; that is, $\outsideoption=(0,0)$. Payoffs are as follows. If the final allocation is $\{(q_t,\transfer_t)\}_{t\in\{1,2\}}$, the buyer and the seller's payoffs are $U(\cdot,\type)=\sum_{\periodone}^\terminal\delta^{t-1}\left(q_t\type-\transfer_t\right)$ and $W(\cdot,\type)=\sum_{\periodone}^\terminal\delta^{t-1}\transfer_t$, respectively, where $\delta\in(0,1)$ is a common discount factor.
\end{ex}

\subsection{Canonical mechanisms and assessments}
\autoref{theorem:main-theorem}  singles out one mechanism-selection game and a class of assessments that allows us to replicate any equilibrium outcome of \gamecollection, for any collection \collection\ of input and output messages. We dub this extensive form the canonical game and the class of assessments, canonical assessments, which we formally define next.
%

\textbf{Canonical game:} \label{page-can-game}The canonical game is the mechanism-selection game in which $\collection=\{(\Types,\Posteriors)\}$. 
We denote the set of equilibrium outcomes of the canonical game by \eqbmoutcomes. 
\begin{definition}[Canonical Mechanisms]\label{definition:canonical-mechanisms} A mechanism $(\Types,\Posteriors,\spot)$ is \emph{canonical} if the mapping $\spot:\Types\mapsto\Delta(\Posteriors\times\allocations)$ can be obtained as the \emph{composition} of two mappings, $\beta:\Types\mapsto\Delta(\Posteriors)$ and $\alpha:\Posteriors\mapsto\Delta(\allocations)$. Formally, for each $\type$ and each pair of measurable subsets, $\measurablem\subset\Posteriors$ and $\measurablea\subset\allocations$, $\spot(\measurablem\times\measurablea|\type)=\int_{\measurablem}\alpha(\measurablea|\posterior)\beta(d\posterior|\type)$.
\end{definition}
In a canonical mechanism, conditional on the output message, the allocation is drawn \emph{independently} of the agent's type report. We refer to the mappings $\beta$ and $\alpha$ as the mechanism's disclosure and allocation rules, respectively. Interpreted as a statistical experiment, $\beta$ encodes how much information the principal learns about the agent's type. Instead, $\alpha$ describes the mechanism's (possibly randomized) allocation, given the information that the principal learns about the agent's type. We let \canonical\ denote the set of canonical mechanisms.

\begin{rem}[Comparison with direct revelation mechanisms]\label{rem:rp} 
A direct revelation mechanism is a special case of a canonical mechanism. To see this, recall that a direct revelation mechanism is a map $\tilde{\alpha}$ assigning to each type \type\  a distribution over allocations; that is, $\tilde{\alpha}:\Types\mapsto\Delta(\allocations)$. A direct revelation mechanism then corresponds to a canonical mechanism $(\beta,\alpha)$, where $\beta$ assigns $\type$ with probability $1$ to the Dirac measure on $\type$, $\delta_\type$, and then sets  $\alpha(\cdot|\delta_\type)=\tilde{\alpha}(\type)$.
\end{rem}

\textbf{Canonical assessments:} A canonical assessment specifies behavior for the principal and the agent that is, in a sense, simple. First, the principal always chooses canonical mechanisms. Second, the agent best responds to the principal's equilibrium choice of mechanisms by participating. Third, input and output messages have \emph{literal} meaning: Conditional on participating, the agent truthfully reports her type, and if the mechanism outputs $\posterior\in\Posteriors$, \posterior\ coincides with the principal's updated beliefs about the agent's type. Formally:

\begin{definition}[Canonical assessments]\label{definition:canonical-assessment}
An assessment \assessment\ of mechanism-selection game \gamecollection\ is canonical if the following holds for all $\periodgeqone$ and all public histories \publict:
\begin{enumerate}
\item\label{itm:principal} The principal offers canonical mechanisms, that is, $\stratpt(\publict)(\canonical)=1$.
\item For all mechanisms $\mechanismt$ in the support of the principal's strategy at \publict, 
\begin{enumerate}[leftmargin=*]
\item\label{itm:participate} For all types \type\ in the support of the principal's beliefs in period $t$, $\beliefend_t(\publict)$, $\participate_t(\type,\hagt,\mechanismt)=1$,
\item\label{itm:truth} For all types \type\ in \Types, $\comm_t(\type,\hagt,\mechanismt)=\delta_\type$, and
\item The mechanism's output belief \posterior\ coincides with the principal's updated belief about the agent's type. Formally, for $\publictplus=(\publict,\mechanismt,1,\posterior,\cdot)$, the marginal of $\beliefend_{t+1}(\publictplus)$ on \Types, $\beliefend_{t+1\Types}(\publictplus)$, coincides with $\posterior$.\footnote{The $1$ in $(\publict,\mechanismt,1,\posterior,\cdot)$ denotes the agent's decision to participate in mechanism \mechanismt.}
\end{enumerate}
%
\item\label{itm:public} The agent's strategy depends only on her private type and the public history.\footnote{Whereas items \ref{itm:participate} and \ref{itm:truth} of \autoref{definition:canonical-assessment} imply the agent's strategy depends only on her private type and the public history \emph{on the path} of the equilibrium strategy starting at \publict, \autoref{itm:public} implies this property also holds \emph{off the path} of the equilibrium strategy starting at \publict, e.g., when the principal deviates and offers a mechanism not in the support of $\stratpt(\publict)$.\label{ftn:referee-3-public-pbe}}
\end{enumerate}
\end{definition}

We let $\ceqbmoutcomes$ denote the set of equilibrium outcomes of the canonical game that are induced by canonical PBE assessments (henceforth, canonical PBE). 
\subsection{Discussion}\label{sec:discussion}
We now discuss three aspects of the model that are important for what follows: the principal may offer randomized allocations, the principal and the agent have access to public randomization, and output messages are public.

\textbf{Randomized allocations:} 
%
Randomized allocations are necessary to conclude that without loss of generality output messages coincide with the principal's posterior beliefs about the agent's type. Indeed, the principal could use \outputt\ to encode randomizations on the allocation; for example, two tuples, $(s_t,a_t)$ and $(s_t^\prime,a_t^\prime)$, may be associated with the same posterior belief. Because a canonical mechanism allows the principal to randomize on the allocation conditional on the posterior belief, we can collapse $\signalt$ and $\signalt^\prime$ to one output message (the posterior belief).

\textbf{Public randomization:} Public randomization allows us to subsume two ways in which the principal may use the mechanism to coordinate continuation play in a PBE of the mechanism-selection game that would otherwise not be possible in a canonical PBE.

First, in the mechanism-selection game, the principal could use \outputt\ to coordinate continuation play;
for example, two tuples $(s_t,\actiont),(s_t^\prime,\actiont)$ may be associated with two different continuation equilibria, but with the same posterior belief. This is one place where the requirement that in a canonical PBE output messages must coincide with the principal's updated beliefs may  be more restrictive than allowing for arbitrary PBE assessments: beliefs are not a rich enough language to encode both the principal's updated beliefs and the suggested continuation play.\footnote{Note this is not a matter of cardinality, but a consequence of the restriction that output beliefs must coincide with equilibrium beliefs. Ultimately, by Kuratowski's theorem (\citealp{srivastava2008course}), $\Posteriors$ is in bijection with $\Posteriors\times[0,1]$, so the number of messages is sufficient to encode both the beliefs and the suggested continuation play.} As we explain after the statement of \autoref{theorem:main-theorem}, the public randomization device in the canonical game allows us to subsume this \emph{coordination} role of the output messages, which, in turn, allows us to conclude that without loss of generality, output messages coincide with the principal's posterior beliefs about the agent's type.\footnote{The same idea arises in the literature on Bayesian persuasion. Implicit in the result in \cite{kamenica2011bayesian} that any experiment can be written as a distribution over posteriors is the assumption that the receiver breaks ties in favor of the sender. Unlike in Bayesian persuasion, it is not clear that the players may be indifferent between two continuation equilibria, so the public randomization device does not generally reduce to simple tie-breaking.\label{ftn:tie-breaking}}  

Second, in the mechanism-selection game, the principal could use the ``name'' of the mechanism to coordinate continuation play. To be concrete, consider \autoref{fig:mix-ms}: In \periodone, the principal randomizes between two mechanisms, \mechanism\ and \mechanismb, each of which is followed by different continuation play, denoted by continuation(\mechanism) and continuation(\mechanismb), respectively. To replicate this in a canonical PBE, we need to determine for each mechanism the distribution over \periodone-allocations induced by the mechanism together with the agent's strategy.  Suppose when coupled with the agent's strategy both \mechanism\ and \mechanismb\ lead to the \emph{same} canonical mechanism, $\mechanism^C$. This is another way in which the language of canonical mechanisms is \emph{coarser} than that of the mechanisms in \mechanismscollection. In the mechanism-selection game, different indirect mechanisms can lead to different continuation equilibria, but to the same canonical mechanism.
%
%
%
%
It is natural to conclude that, similar to the use of the public randomization device in the previous paragraph, we could replicate play in the mechanism-selection game via a canonical PBE in the canonical game as illustrated in \autoref{fig:canonical-cor}. In the canonical game, 
%
the principal offers mechanism $\mechanism^C$ in \periodone, and then we use the public randomization device in \periodtwo\ to replicate the continuation play associated with \mechanism\ and \mechanismb\ in the original assessment.
\begin{figure}[h!]
\centering
\subfloat[Mechanism-selection game]{
\begin{tikzpicture}[thick,scale=0.5]
\node[nsnode,label=left:{}](origin) at (0,0){};
\node[bsnode,label=above:{\mechanism\ }](m1) at (2,1){};
\node[bsnode,label=below:{\mechanismb\ }](m2) at (2,-1){};
\node[label=right:{continuation (\mechanism)}](m1c) at (4,1){};
\node[label=right:{continuation (\mechanismb)}](m2c) at (4,-1){};

\draw[black](origin) edge node[above left]{$\frac{1}{2}$} (m1);
\draw[black](origin) edge node[below left]{$\frac{1}{2}$} (m2);
\draw[black](m1) edge node [above]{} (m1c);
\draw[black](m2) edge node [above]{} (m2c);
\end{tikzpicture}\label{fig:mix-ms}
}

\subfloat[Canonical game: \periodtwo-public randomization]{\begin{tikzpicture}[thick,scale=0.5]
\node[nsnode,label=left:{}](origin) at (0,0){};
\node[bsnode,label=above:{$\mechanism^C$}](cm1) at (2,0){};
\node[label=right:{$\cor_2$, continuation (\mechanism)}](m1c) at (5,1){};
\node[label=right:{$\corb_2$, continuation (\mechanismb)}](m2c) at (5,-1){};

\draw[black](origin) edge node[above]{} (cm1);
\draw[black](cm1) edge node [above left]{$\frac{1}{2}$} (m1c);
\draw[black](cm1) edge node [below left]{$\frac{1}{2}$} (m2c);
\end{tikzpicture}\label{fig:canonical-cor}}
\hspace{1cm}
\subfloat[Canonical game: \periodone-public randomization]{\begin{tikzpicture}[thick,scale=0.5]
\node[nsnode,label=left:{}](origin) at (0,0){};
\node[bsnode,label=above:{$\mechanism^C$ }](m1) at (3,1){};
\node[bsnode,label=below:{$\mechanism^C$ }](m2) at (3,-1){};
\node[label=right:{ continuation (\mechanism)}](m1c) at (5,1){};
\node[label=right:{ continuation (\mechanismb)}](m2c) at (5,-1){};

\draw[black](origin) edge node[above left]{$\cor_1,\frac{1}{2}$} (m1);
\draw[black](origin) edge node[below left]{$\corb_1,\frac{1}{2}$} (m2);
\draw[black](m1) edge node [above left]{} (m1c);
\draw[black](m2) edge node [below left]{} (m2c);
\end{tikzpicture}\label{fig:cor-canonical}
}
\caption{Different mechanisms lead to the same canonical mechanism}\label{fig:cor-2}
\end{figure}

However, the construction in \autoref{fig:canonical-cor} may not be enough to replicate the original outcome distribution. To see this, suppose in the mechanism-selection game, both \mechanism\ and \mechanismb\ are accepted with probability $1$, so the principal's beliefs after non-participation are not pinned down by Bayes' rule. Furthermore, assume different off-path beliefs and different continuation equilibria are associated with the rejection of \mechanism\ and \mechanismb. However, we can only assign \emph{one} belief and \emph{one} continuation equilibrium to the event in which the agent rejects $\mechanism^C$ in \autoref{fig:canonical-cor}. It may not be possible to find one off-path belief and one continuation equilibrium that simultaneously make it optimal for the agent to participate in $\mechanism^C$ and sequentially rational for the principal to follow the prescribed continuation play.
%

Instead, we can replicate the original outcome distribution using the construction in \autoref{fig:cor-canonical}: We use the public randomization device in \periodone\ to encode the indirect mechanism that led to $\mechanism^C$ in the mechanism-selection game. It follows that in order to replicate the outcome distributions via canonical PBE public randomization may be needed even before play begins.

While the proof of \autoref{theorem:main-theorem} deals explicitly with the coordination role of the output message, it does not deal explicitly with the the issue illustrated in \autoref{fig:cor-2}, which can only arise if the principal is using mixed strategies. Instead, we first show in Lemma D.1 that it is without loss of generality to assume that the principal plays a pure strategy in the mechanism-selection game. Similar to the construction in \autoref{fig:cor-canonical}, we use the public randomization device in the mechanism-selection game to encode the mechanisms over which the principal is randomizing, effectively purifying the principal's strategy.
%

%

\textbf{Public output messages:} \label{page-public}Because in the mechanism-selection game the output message is public, the agent is the only player with private information, which is the leading informational setting in the literature on mechanism design with limited commitment and short-term contracts referenced in the introduction. Indeed, our only point of departure from this literature is the class of mechanisms we endow the principal with. 

If, instead, output messages were observed only by the principal, he would become \emph{endogenously} privately informed. Having an informed principal adds a new friction even in a one-shot interaction: the mere choice of a mechanism serves as a signal of the principal's private information. Little is known about dynamic and \emph{exogenously} informed principal problems under commitment, let alone under limited commitment. For these reasons, the comparison of the equilibrium outcomes of the mechanism-selection game with public and private output messages is an open question.


%

\section{Revelation principle}\label{sec:main}

\autoref{sec:main} presents the main result of the paper: To characterize the set of equilibrium outcomes that can arise in some mechanism-selection game, it is enough to characterize the canonical PBE outcomes of the canonical game. Formally,

\begin{theorem}[Revelation principle]\label{theorem:main-theorem}
For any PBE outcome of any mechanism-selection game \gamecollection, an outcome-equivalent canonical PBE of the canonical game exists. That is,
\begin{align*}
\bigcup_\collection\eqbmoutcomes_\collection=\eqbmoutcomes=\ceqbmoutcomes.
\end{align*}
%
\end{theorem}

\autoref{theorem:main-theorem} plays the same role in mechanism design with limited commitment as the revelation principle does in the commitment case. First, it identifies a well-defined set of mechanisms, \canonical, that, without loss of generality, the principal uses to implement any equilibrium outcome. 
%
Second, it simplifies the analysis of the behavior of the agent in the game induced by the mechanisms chosen by the principal: we can always restrict attention to assessments in which the agent participates and truthfully reports her type. As we illustrate through the application in \autoref{ex:sdg}, this restriction allows us to reduce the agent's behavior to a set of constraints that the mechanism must satisfy, as in the case of commitment.

The proof of \autoref{theorem:main-theorem} follows from two observations. The first one is easy: because the canonical game is a mechanism-selection game, any PBE outcome of the canonical game is a PBE outcome of some mechanism-selection game; that is, $\eqbmoutcomes\subseteq\cup_\collection\eqbmoutcomes_\collection$. The second one constitutes the bulk of the proof, which we overview below: we show that for any collection \collection\ and for any PBE assessment of the mechanism-selection game \gamecollection, an outcome-equivalent canonical PBE assessment of \gamecollection\ exists. Because in the canonical game the principal has fewer deviations than in \gamecollection\ \emph{and} in a canonical PBE assessment the principal plays a strategy that is available in the canonical game, it follows that for any PBE assessment of the mechanism-selection game, an outcome-equivalent canonical PBE assessment of the canonical game exists; that is $\eqbmoutcomes_\collection\subseteq\ceqbmoutcomes$. Because $\ceqbmoutcomes\subseteq\eqbmoutcomes$, this concludes the proof. 

We now review the steps involved in the proof that any PBE outcome of the mechanism-selection game \gamecollection\ can be achieved in a canonical PBE of the canonical game.\footnote{\autoref{appendix:propositions} proves \autoref{theorem:main-theorem} under the assumption of finitely many types and that the principal only offers mechanisms \mechanism\ such that for all input messages $m\in\inputm$, the support of $\varphim(\cdot|m)$ is finite. 
This simple proof mirrors the complete one in Online Appendix D.1, but it is technically simpler and more accessible. 
} To simplify the presentation, we rely on the following construction.\footnote{This construction is the standard one whereby the non-participation decision is included as an option in the mechanism. At the end of the overview, we address why we model participation as a separate decision.} Given a mechanism \mechanism\ and the agent's participation and reporting strategies $(\participate,\comm)$, we can extend the principal's mechanism and the agent's reporting strategy as follows. Extend the set of input messages so as to include a non-participation message, $\inputm_\nmssg=\inputm\cup\{\emptyset\}$. Similarly, extend the set of output messages, $\outputm_\emptyset=\outputm\cup\{\nsignal\}$ and define $\mechanism_\participate=(\inputm_\nmssg,\outputm_\emptyset,\varphim_\participate)$ as follows: $\varphim_\participate$ coincides with \varphim\ on \inputm\ and assigns probability $1$ to $\{(\nsignal,\outsideoption)\}$ for $m=\nmssg$. Finally, define $\comm_\participate$ as follows: $\comm_\participate$ sends message \nmssg\ with probability 1-\participate, and sends message $m\in\inputm$ with probability $\participate\comm(m)$.
%

\textbf{Input messages as type reports:} To fix ideas, consider the proof for the standard revelation principle in static settings. The extended mechanism, $\mechanism_\participate$, together with the agent's extended reporting strategy, induce a mapping from \Types\ to distributions over $\outputm_\nsignal\times\allocations$. This allows us to conclude we can replace the set of input messages with the set of type reports, as illustrated in \autoref{fig:static-rp}.
\begin{figure}[h!]
\begin{minipage}{0.5\textwidth}
\begin{center}
\subfloat[Static revelation principle]{
\begin{tikzpicture}
\node[label=left:{$\Types$}](types) at (0,0){};
\node[label=right:{$\outputm_{\nsignal}\times\allocations$}](output) at (3,0){};
\node[label=below:{$\inputm_{\nmssg}$}](input) at (1.5,-1){};
\draw[->](types)edge node[below left]{$\comm_\participate$}(input);
\draw[->](input)edge node[below right]{$\varphim_\participate$}(output);
\draw[->,dashed](types) edge node[above]{$\varphim_\participate\circ r_\participate$}(output);
\end{tikzpicture}\label{fig:static-rp}}
\end{center}
\end{minipage}
\begin{minipage}{0.5\textwidth}
\begin{center}
\subfloat[Mechanism-selection game]{
\begin{tikzpicture}
\node[label=left:{$\Types\times\Hagt(\publict)$}](types) at (0,0){};
\node[label=right:{$\outputt_{\nsignal}\times\allocations$}](output) at (3,0){};
\node[label=below:{$\inputt_{\nmssg}$}](input) at (1.5,-1){};
\draw[->](types)edge node[below left]{$\comm_\participate$}(input);
\draw[->](input)edge node[below right]{$\varphit_\participate$}(output);
\draw[->,dashed](types) edge node[above]{$\varphit_\participate\circ r_\participate$}(output);
\end{tikzpicture}\label{fig:mech-selection-rp}}
\end{center}
\end{minipage}
\caption{Type reports as input messages}
\end{figure}

In the dynamic setting, however, this argument would only allow us to conclude we can rewrite the mechanism as a mapping from $\Types\times\Hagt(\publict)$ to distributions over $\outputt_\nsignal\times\allocations$, as illustrated in \autoref{fig:mech-selection-rp}. Indeed, to replicate the agent's reporting strategy, the mechanism needs to obtain all the information on which the agent conditions her strategy, which potentially is $(\type,\hagt)$.

However, we show that given a PBE in which the agent conditions her strategy on the payoff-irrelevant part of her private history at some public history \publict, another outcome-equivalent PBE exists in which she does not 
(see \autoref{prop:fp-private-history-irrelevance}).\footnote{This result is useful also in applications. It states that in our game, the set of PBE payoffs coincides with the set of public PBE payoffs (\citealp{athey2008collusion}). In games with time-separable payoffs, public PBE payoffs are amenable to self-generation techniques, as in \cite{abreu1990toward}. See \cite{doval2020optimal}.} Thus, conditional on the public history \publict, the mechanism, together with the agent's reporting strategy, induces a mapping from \Types\ to distributions over $\outputt_\nsignal\times\allocations$, so we can always take the set of input messages to be the set of type reports. This result relies on two observations. First, because input messages are payoff irrelevant and unobserved by the principal, if the agent chooses different strategies at $(\type,\hagt)$ and $(\type,\hagtc)$ with $\hagt, \hagtc\in\Hagt(\publict)$, she is indifferent between these two strategies. However, the principal may not be indifferent between these two strategies. Second, we build an alternative strategy for the agent that conditions only on $(\type,\publict)$ and yields the same outcome distribution, and hence the same payoff for the principal.

Two implications follow from this step. First, given a public history \publict\ and the principal's choice of mechanism at \publict, \mechanismt,  the \emph{auxiliary} mapping, $\varphit_\participate\circ\comm_\participate$, only takes $\Types$ as an input. The mapping $\varphit_\participate\circ\comm_\participate$ is a \emph{direct} mechanism, where the set of input messages are type reports, and for which the agent finds it optimal to truthfully report her type. Second, the relevant part of the principal's beliefs in the game are about the agent's type, \type, and not the agent's private history, \hagt. This finding is important because although the auxiliary mapping allows us to replicate the distribution over period-$t$ outcomes induced by the agent's strategy and the mechanism \mechanismt, in a dynamic game, we also need to replicate the distribution over continuation play, and the principal's beliefs are an important component of continuation play. 
%
%

\textbf{Output messages as beliefs and canonical mechanisms:} \label{page-output}As discussed in \autoref{sec:discussion}, the principal may have other uses for the output messages beyond encoding information about the agent: Conditional on the same posterior belief, (i) he may encode randomizations over the allocation, and (ii) he may use the output message to coordinate continuation play. As the proof of \autoref{theorem:main-theorem} shows, randomized allocations and public randomization allow us to subsume these two roles of the output messages.

Now, the potential challenge in using the public randomization device to subsume the second role of the output message is that, by definition, the use of the public randomization device in the canonical game can only depend on publicly available information. Instead, because in the auxiliary mapping the agent is reporting truthfully, the output message in the mechanism-selection game is drawn as a function of the agent's type, which allows the principal to coordinate future play above and beyond what he would be able to do by solely relying on the public randomization device in the canonical game.
%

To overcome this challenge, we leverage here that canonical mechanisms use \emph{beliefs} as output messages. Note that beliefs are a sufficient statistic for the information about the agent's type that is encoded in the output messages of the mechanism-selection game. Thus, conditional on the induced belief and the allocation, the selection of continuation play contains no further information about the agent's type, which allows us to \emph{decompose} the mechanism in the mechanism-selection game into a canonical mechanism in the canonical game that uses beliefs as output messages and a public randomization device.

The above argument also explains why, in a canonical mechanism, conditional on the output message, the allocation can be drawn independently of the agent's type (report). Ultimately, conditional on the induced belief, the allocation contains no further information about the agent's type.

Briefly, the proof of this result proceeds as follows (see \autoref{prop:fp-bijection}). Suppose that the principal offers \mechanismt\ in period $t$. The principal's belief about the agent's type, $\beliefend_t(\publict)$, together with the auxiliary mapping, $\varphit_\participate\circ\comm_\participate$, induces a joint distribution over $\Types\times\outputt_\nsignal\times\allocations\times\Posteriors\times\Cor$, where $\Cor=[0,1]$ denotes the public randomization device in the mechanism-selection game. Because conditional on the induced posterior, $(s_t,\actiont)\in\outputt_\emptyset\times\allocations$ carries no further information about the agent's type, this allows us to ``split'' the mechanism into a transition probability $\beta$ from \Types\ to $\Posteriors$, a transition probability $\alpha$ from \Posteriors\ to \allocations, and a transition probability $\omega$ from $\Posteriors\times\allocations$ to $\outputt_{\nsignal}\times\Cor$. The transition probability $\alpha$ plays the first role of the output message and highlights the importance of allowing the principal to offer randomized allocations. The transition probability $\omega$ corresponds to the public randomization device: by Kuratowski's theorem, we can always embed $\outputt_{\nsignal}\times\Cor$ into $[0,1]$ (see \citealp{srivastava2008course}).

Three conceptual insights arise from this result. First, when the mechanism is canonical, the principal can separate the design of the information that the mechanism encodes about the agent's type from the design of the allocation. Second,  the allocation has to be measurable with respect to the information generated by the mechanism: The more the principal desires to tailor the allocation to the agent's type, the more he has to learn about the agent's type through the mechanism.\footnote{Contrast this case with the one in which the principal has commitment, where we write a mechanism as a menu of options, one for each type of the agent. We do so even if the optimal mechanism offers the same allocation to a set of agent types. When the principal has commitment, whether the allocation reveals more information beyond the set of types that receive that allocation is irrelevant, because additional information can always be ignored. Under limited commitment, however, this is not the case, and the principal in general trades off tailoring the allocation to the agent's type and the information that is learned through this.} Third, it highlights the \emph{coordination} role of the mechanism, which is subsumed by the public randomization device: beyond determining today's allocation and the information that is carried forward in the interaction, the mechanism allows the principal to coordinate future play.
%

\textbf{Bayes' rule, truthtelling, and participation:} Underlying the previous step is the assumption that the beliefs associated with the output messages are determined via Bayes' rule. In particular, the principal is never surprised by any output message he observes. To ensure that beliefs are pinned down by Bayes' rule, \autoref{prop:fp-bijection} shows we can ``eliminate'' from the mechanism all input messages that are used only by types to whom the principal assigns $0$ probability. Eliminating these input messages, however, may change the participation decision for types not in the support of the principal's beliefs, which is why a canonical PBE assessment does not require that these types participate in the mechanism.\footnote{If an agent's type has zero probability at a specific public history, she can only reach it through a deviation from \strat. This change to the mechanism actually makes the deviation less attractive, and hence, it ``disincentivizes" the agent from deviating in the first place.} 

Instead, it should be intuitive that the agent participates in the mechanism whenever her type is in the support of the principal's beliefs: By relying on the map between output messages and posterior beliefs and the public randomization device, we guarantee that, conditional on participation, the agent faces the same 
period-$t$ allocation and distribution over continuations as when she did not participate. The map between output messages and beliefs allows us to identify which output message one should associate with the types that chose not to participate: the one that corresponds to the principal's updated belief conditional on non-participation.\footnote{Starting from an equilibrium in which the mechanism is rejected with positive probability, this belief is also determined via Bayes' rule.} The map between output messages and the public randomization device allows us to replicate the distribution over continuations the agent faces in the PBE of the mechanism-selection game for those types that found it optimal to randomize between participating and not participating. 

That beliefs in the support of the mechanism are determined via Bayes' rule has one practical implication that we exploit throughout our analysis of \autoref{ex:sdg} and in our concurrent work (\citealp{doval2020optimal,doval2021purchase}): the mechanism's disclosure rule together with the principal's belief about the agent's type induce a Bayes' plausible distribution over posteriors. As a consequence, we can apply tools from information design to derive qualitative properties of the principal's problem (see \autoref{prop:finite-support}). This is where modeling participation as a decision that happens outside the mechanism as opposed to as an input message that locks the outside option as in \autoref{fig:mech-selection-rp} is important: because non-participation is a zero-probability event, the principal's beliefs after non-participation are not pinned down via Bayes' rule, and hence cannot be necessarily obtained from a Bayes' plausible distribution over posteriors. At the same time, these beliefs cannot be ignored in the analysis, because they determine the continuations after non-participation, and hence, the agent's incentives to participate in the first place. By modeling participation as a decision that happens outside of the mechanism, we can rely on the tools of information design to design the mechanism's disclosure rule, while, as we illustrate using the application in \autoref{ex:sdg}, the participation decision is summarized by a participation constraint (see \ref{eq:seller-opt}).

\subsection{The revelation principle at work}\label{sec:sdg-program}
We now illustrate the simplifications afforded by \autoref{theorem:main-theorem} within the context of the application in \autoref{ex:sdg}. In particular, we show that in order to characterize the seller's revenue maximizing PBE outcome it is enough to characterize the solution to a constrained optimization problem, denoted \ref{eq:seller-opt}, that only involves the seller. This is already in stark contrast to the existing work in mechanism design with limited commitment, which needs to keep track of how the buyer's best response to the seller's mechanism determines the information that the seller obtains from the interaction, which, in turn, affects the seller's incentives to offer the mechanism in the first place.


To arrive at the program that characterizes the seller's maximum revenue, we appeal to \autoref{theorem:main-theorem}. First, in what follows, we restrict attention to the canonical game and to assessments in which the seller offers canonical mechanisms. Second, without loss of generality, we can consider assessments where the buyer's strategy does not depend on the payoff-irrelevant part of the private history. In particular, in \periodtwo, the seller's optimal mechanism only needs to elicit the buyer's payoff relevant type, \type. Let $\posterior_2$ denote the seller's \emph{posterior} belief in \periodtwo. The optimal mechanism in \periodtwo\ is a posted price \emph{regardless} of the properties of $\posterior_2$ (see Proposition 2 in \citealp{skreta2006sequentially}). For each belief the seller may have in \periodtwo, $\posterior_2$, we let $p_2(\posterior_2)$ denote a selection from the set of optimal prices in \periodtwo\ when his belief is $\posterior_2$.\footnote{The seller may be indifferent in \periodtwo\ among several prices. In that case, as in the literature on Bayesian persuasion, we determine the tie-breaking rule as a solution to the seller's problem in \periodone\ (see \ref{eq:seller-opt}). Note that because of public randomization the selection from the set of optimal prices in \periodtwo\ at belief $\posterior_2$ may be random. However, our notation does not account for this explicitly  to simplify the presentation.}

Third, it is without loss of generality to consider assessments in which (i) the buyer's best response to the seller's equilibrium choice of mechanism in \periodone\ is to participate and truthfully report her type with probability $1$, and (ii) when the output message is $\posterior_2$, the seller updates his belief to $\posterior_2$.  Moreover, the assumption of quasilinearity implies that, without loss of generality, the seller does not randomize on the transfers: below $\transfer(\posterior_2)$, denotes the expected payment conditional on $\posterior_2$, and $q(\posterior_2)\in[0,1]$ denotes the probability with which the good is sold. Thus, we can write the seller's problem as follows:
\begin{align}\label{eq:seller-opt}\tag{OPT}
&&&\max_{(q,\beta,\transfer,p_2)}\mathbb{E}_{\prior}\left[\int_{\Posteriors}\left(\transfer(\posterior_2)+(1-q(\posterior_2))\delta p_2(\posterior_2)\mathbbm{1}[\type\geq p_2(\posterior_2)]\right)\beta(d\posterior_2|\type)\right]\\
\text{s.t.}&&&
(\forall\type\in\Types)U(\type)=\int_{\Posteriors}\left(\type q(\posterior_2)-\transfer(\posterior_2)+(1-q(\posterior_2))\delta u^*(\type,\posterior_2)\right)\beta(d\posterior_2|\type)\geq0\label{eq:sdg-pc}\tag{PC}\\
&&&(\forall\type,\typed\in\Types)U(\type)\geq\int_{\Posteriors}\left(\type q(\posterior_2)-\transfer(\posterior_2)+(1-q(\posterior_2))\delta u^*(\type,\posterior_2)\right)\beta(d\posterior_2|\typed)\label{eq:sdg-ic}\tag{IC}\\
&&&(\forall\type\in\Types)(\forall\measurablem\subset\Posteriors)\int_{\measurablem}\beliefend_2(\type)\left(\mathbb{E}_{\typec}\left[\beta(d\posterior_2|\typec)\prior(\typec)\right]\right)=\beta(\measurablem|\type)\prior(\type).\label{eq:sdg-bp}\tag{BP}
\end{align}
That the seller's belief about the buyer's type updates to $\posterior_2$ when the output message is $\posterior_2$ appears twice in the above expression: First, in \autoref{eq:sdg-bp}, which is the Bayes' plausibility constraint and, second, in the objective function, where the seller's payoff in \periodtwo\ when the agent's type is \type\ and his belief is $\posterior_2$ corresponds to whether \type\ buys the good at a price of $p_2(\posterior_2)$. The latter affords an important simplification: instead of writing the seller's program as one in which the seller chooses a mechanism for period $1$ and one for period $2$ subject to the constraint that the period $2$ mechanism is optimal given the seller's belief in \periodtwo, the program \ref{eq:seller-opt} has the seller maximize over  one-period mechanisms by replacing the seller's best response in \periodtwo\ in the seller's objective function.

The two remaining constraints in \ref{eq:seller-opt} are the buyer's participation and incentive compatibility constraints (Equations \ref{eq:sdg-pc} and \ref{eq:sdg-ic}). The buyer's payoff in the mechanism, $U(\type)$, is determined as follows. For each $\posterior_2$ in the support of $\beta(\cdot|\type)$, she receives the good with probability $q(\posterior_2)$ and makes a payment of $\transfer(\posterior_2)$; with the remaining probability, no trade occurs, and she obtains a continuation payoff, $u^*(\type,\posterior_2)$, which describes her optimal decision of whether to buy the good at $p_2(\posterior_2)$. The participation constraint states that the buyer has to earn a payoff of at least $0$ by participating. Indeed, because non-participation is a $0$ probability event, we can specify that upon rejection of the mechanism, the seller believes the buyer's valuation is \maxt, so that in $\periodone$, the seller chooses a price of $\maxt$ when the buyer chooses not to participate. The incentive compatibility constraint states that when her type is $\type$, the buyer cannot obtain a higher payoff by reporting that her type is $\typed\neq\type$. When the buyer reports \typed, she obtains a different distribution over output messages $\beta(\cdot|\typed)$; however, in \periodtwo, she still chooses optimally whether to buy the good, which explains the term $u^*(\type,\posterior_2)$. 

The three constraints in \ref{eq:seller-opt} provide us with a tractable representation of both the buyer's behavior and its impact on the mechanism offered in \periodtwo\ via the information that is generated about the buyer's type in \periodone. Thus, instead of having to consider complicated mixed strategies on the part of the agent (see \citealp{laffont1988dynamics,bester2001contracting}), we have reduced the problem of characterizing the seller-optimal PBE outcome to the solution of a program \ref{eq:seller-opt} that combines elements of information design and mechanism design. Indeed, the solution to \ref{eq:seller-opt} can be leveraged 
to fully specify the PBE assessment that implements the seller's maximum revenue.\footnote{For an illustration of this, see our companion work, \cite{doval2020optimal}.} 

%
Furthermore, as we show below in \autoref{prop:finite-support}, when \Types\ is finite, \ref{eq:seller-opt} can be further simplified: without loss of generality we can assume the seller employs mechanisms such that the support of $\beta$ is finite for all $\type\in\Types$. 
\begin{prop}\label{prop:finite-support}
Suppose \Types\ is finite. Fix \collection\ and let \assessment\ denote a canonical PBE assessment of \gamecollection. Then, a payoff-equivalent canonical PBE assessment exists such that for all \periodgeqone\ and all $\publict$, the principal's choice of mechanism at \publict, \mechanismt, satisfies that for all $\type\in\Types$, the support of $\betat(\cdot|\type)$ is finite.
\end{prop}
The proof of \autoref{prop:finite-support} highlights that \autoref{eq:sdg-bp} implies that the mechanism's disclosure rule, $\beta$, induces a Bayes' plausible distribution over posteriors. 
Like in the literature on Bayesian persuasion, we can then rely on Carath\'eodory's theorem (\citealp{rockafellar2015convex}) to ensure that the principal and the agent's payoffs remain the same should the principal use a mechanism that employs finitely many posteriors. 

Most of the existing analysis of the model in \autoref{ex:sdg} is performed for continuum type spaces; we thus revisit \ref{eq:seller-opt} when \Types\ is a continuum in \autoref{sec:sdg}. Before doing so, we first explain why it is impossible to endow the set of mechanisms with a measure structure so that the game is well-defined and, then, develop a framework  in \autoref{sec:dynamic-mechanisms} that is suitable to study mechanism design under limited commitment with continuum types spaces.

\section{Continuum type spaces}\label{sec:continuum}
\autoref{sec:continuum} considers the case in which the set of types is an uncountable compact Polish space. This extension is important because much of the standard toolkit of mechanism design has been developed for continuum (and convex) type spaces, where the representation of incentive compatible mechanisms can be obtained using the envelope theorem. \autoref{sec:complications} reviews the issues raised by \cite{aumann1961borel}, and hence the difficulties with having a well-defined mechanism-selection game when \Types\ is uncountable. In particular, we explain why the usual solution to this problem, namely, restricting the principal to choosing mechanisms in a suitably defined set, is not enough for the purpose of deriving a revelation principle. With little loss of continuity, the reader can skip to \autoref{sec:dynamic-mechanisms}, where we propose a framework that allows us to sidestep the aforementioned issues and obtain the analogue of \autoref{theorem:main-theorem} when \Types\ is uncountable. We then apply our results to \autoref{ex:sdg}.

\subsection{Choosing functions at random}\label{sec:complications}
To define the mechanism-selection game, a measurable structure on \mechanismscollection\ is needed to define (i)  the principal’s mixed strategies, (ii) the principal and the agent’s expected payoffs from those mixed strategies, and (iii) the principal and the agent’s strategies as measurable functions of the histories, which include the past choices of the mechanisms. Focusing on (i) and (ii), \citet[Theorem D]{aumann1961borel} implies no suitable measure structure on \mechanismscollection\ exists when the set of input messages is uncountable. Instead, \cite{aumann1964mixed} circumvents the issue of defining a measurable structure on the set \mechanismscollection\ to define mixed strategies by relying on randomization devices. However, the construction in \cite{aumann1964mixed} is insufficient for our purposes because the mechanisms chosen by the principal through period $t-1$ are part of the public histories.
 Thus, to define the principal and the agent's strategies as measurable functions of the histories, we again face the issue of defining a measurable structure on the set of mechanisms and with the negative answers in \cite{aumann1961borel}.

For this reason, the literature on competing principals (see, e.g. \citealp{attar2021keeping,attar2021competing}) follows a different approach: Theorem D in \cite{aumann1961borel} implies that the issues raised above would be mute if we restrict the principal to choosing mechanisms from a subset \Mechanismsb\ of \mechanismscollection, such that \Mechanismsb\ is of bounded Borel class.\footnote{For uncountable \messages, $\{\spot:\messages\mapsto\Delta(S\times A):\spot \text{ is measurable}\}$ is not of bounded Borel class  (\citealp{aumann1961borel}).}  For the purposes of deriving a revelation principle, this approach is again insufficient: Borel classes are not always closed under composition (\citealp{srivastava2008course}) and we obtain a canonical mechanism by composing the agent's strategy with the mechanism the principal employs in the mechanism-selection game. Unless the agent's strategy is continuous in her type, the induced canonical mechanism may be of a Borel class strictly larger than that of the original mechanism. Furthermore, different equilibria of the mechanism-selection game may necessitate canonical mechanisms of different Borel classes, which makes the task of defining the canonical game pointless: one would have to potentially consider a different canonical game for each equilibrium of each mechanism-selection game. Lastly, the restriction to a set of mechanisms of bounded Borel class is difficult to work with in applications: only payoffs from deviations to mechanisms within that class are well-defined and, in practice, verifying that only deviations to those mechanisms are being contemplated is difficult.

Motivated by the importance of continuum type spaces, \autoref{sec:dynamic-mechanisms} proposes an approach to model mechanism-selection games that circumvents the above issues.
\subsection{PBE-feasible outcomes}\label{sec:dynamic-mechanisms}
We develop a framework to characterize the outcomes that can be sustained under limited commitment, which we dub PBE-feasible outcomes and formally define below (\autoref{definition:pbe-feasible}). By analogy with the mechanism-selection game, we keep the notation $\eqbmoutcomes_\collection$ to denote PBE-feasible outcomes. Contrary to the mechanism-selection game, $\eqbmoutcomes_\collection$ is now a correspondence describing the set of PBE-feasible outcomes for each period \periodgeqone, each principal's belief \priort\ and each sequence of allocations up to period $t$, \actionuptot, $\eqbmoutcomes_\collection(\priort,\actionuptot)$. The reason is that the definition of the set of PBE-feasible outcomes is recursive, and what is PBE-feasible in period $t$ naturally depends on what is PBE-feasible in period $t+1$.

%
%

The discussion in \autoref{sec:complications} implies that the framework must sidestep the need to define the principal and the agent's ``strategies'' as measurable functions of the principal's past choices of mechanisms. There are (at least) two important roles measurability plays in the mechanism-selection game. First, it allows us to describe the agent's behavior along the path of the principal's strategy, which in turn allows us to evaluate the principal's payoff from a given strategy. Second, it allows us to describe how the agent's behavior changes when the principal deviates from the prescribed strategy, which in turn allows us to evaluate the principal's payoff from a deviation from the prescribed strategy. The comparison between these two payoffs determines whether the principal's strategy is sequentially rational.

Informally, the framework circumvents the aforementioned measurability issues as follows. The starting point of the analysis is that we no longer model the principal as a player. Instead, we describe the analogue of a principal's strategy in the mechanism-selection game via an extensive-form game for the agent, which describes the sequence of mechanisms the agent faces as a function of her participation decisions and the outcomes of the mechanisms (\autoref{definition:dynamic-mechanisms}). Importantly, in this extensive-form game the agent's strategy only depends on her type, her participation decisions, her input messages, and the outcomes of the mechanisms, but not on the mechanisms themselves. The principal's belief about the agent's type together with the agent's strategy define a distribution over the terminal nodes in this extensive form, which allows us to evaluate the principal's payoff from a particular sequence of mechanisms. The final component of the framework is how we define that a particular extensive form is \emph{sequentially rational} for the principal; in other words, that the principal does not wish to revise the mechanisms that define the extensive form at any point in time (\autoref{definition:dm-sequential-rationality}). Here we rely on two ideas. First, to determine whether the principal wishes to revise the mechanisms that define the extensive form, we only need to know the outcome distributions the principal expects he will face in the event of a deviation (\autoref{eq:deviant}). Second, these outcome distributions can be defined without reference to a strategy of the agent that conditions on the sequence of mechanisms that has been offered so far.

We now formally define the set of PBE-feasible outcomes, $\eqbmoutcomes_\collection(\priort,\actionuptot)$, for each period \periodgeqone, and pair $(\priort,\actionuptot)\in\Posteriors\times\allocationsuptot$ (\autoref{definition:pbe-feasible}). Because the definition is recursive, we fix a period \periodgeqone, and a pair $(\priort,\actionuptot)$ throughout. \autoref{definition:pbe-feasible} consists of three components, which we introduce first: (i) the sequence of mechanisms offered by the principal (\autoref{definition:dynamic-mechanisms}), (ii) optimal behavior by the agent within those mechanisms, and (iii) the outcome distributions the principal anticipates upon a deviation (\autoref{eq:deviant}). For simplicity, we assume the principal has one set of input and output messages, that is, $\collection=\{(\messages,\signals)\}$, and we use the shorthand notation $\saemptyb$ to denote the set $(\signals\times\allocations)\cup\{(\nsignal,\outsideoption)\}$.

\textbf{Dynamic mechanisms:} Instead of having the principal be a player in a game, we describe the analogue of the principal's strategy via a \emph{dynamic mechanism}, defined as follows:

\begin{definition}[Dynamic mechanisms]\label{definition:dynamic-mechanisms}
For \periodgeqone\ and $\actionuptot\in\allocationsuptot$, a dynamic mechanism given \actionuptot, \spott, is a sequence of measurable mappings
$\spotindexb:(\saemptyb\times\Cor)^{\dateindexb-t}\times\messages\mapsto\Delta(\signals\times\allocations)$,
such that for all $\dateindexb\geq t$ and all $(s^{\dateindexb-t},\action^{\dateindexb-t},\cor^{\dateindexb-t})$, 
\begin{enumerate}
\item $\spotindexb(s^{\dateindexb-t},\action^{\dateindexb-t},\cor^{\dateindexb-t},\cdot):\messages\mapsto\Delta(\signals\times\allocations)$ is a measurable function, and
\item for all $m\in\messages$, $\spotindexb(s^{\dateindexb-t},\action^{\dateindexb-t},\cor^{\dateindexb-t})\left(\feasibleallocations_t(\actionuptot,\action^{\dateindexb-t})|m\right)=1$.
\end{enumerate}
\end{definition}
When \periodone, a dynamic mechanism describes the mechanism the agent faces in period $1$, $\spot_1$, the mechanism  the agent faces in period $2$ as a function of the agent's participation decision (i.e., whether $(s_1,\action_1)\neq(\nsignal,\outsideoption)$),
and the realization of the public randomization device,  $\spot_2(s_1,\action_1,\cor_2)$, and so on. Consider now $t>1$ and suppose the allocation so far is \actionuptot. Then, we require that the dynamic mechanism only implements allocations that are feasible given \actionuptot.

As we explain next, a dynamic mechanism defines an extensive-form game for the agent:

\textbf{Agent-extensive form:} Given $(\priort,\actionuptot)$, a dynamic mechanism \spott\ defines an extensive-form game for the agent, \extensive(\priort,\actionuptot,\spott), as follows. First, nature draws the agent's type according to \priort. Having observed her type, suppose that in stage $\dateindexb-t$, the public history is $\publicttau=(s^{\dateindexb-t},\action^{\dateindexb-t},\cor^{\dateindexb-t})$. Then,  faced with $\spot_\dateindexb(\publicttau)$, the agent decides whether to participate and, conditional on participating, her reporting strategy. If the agent rejects $\spot_\dateindexb$, the ``output message'' is \nsignal\ and the allocation is \outsideoption. Instead, if she accepts $\spot_\dateindexb(\publicttau)$, she chooses an input message $m\in\messages$ that determines the distribution from which the output message and the allocation are drawn, $\spot_\dateindexb(\publicttau)(\cdot|m)$.  In both cases, we proceed to stage $\dateindexb+1-t$.

In the agent-extensive form \extensive(\priort,\actionuptot,\spott), there are two types of histories. The public history \publicttau\ encodes the agent's participation in the mechanism, the realized output messages and allocations, and  the realizations of the public randomization device. The private histories encode everything the agent knows: her payoff-relevant type \type, the public history \publicttau, and her past input messages. In a slight abuse of notation, we denote by \Privatettau(\publicttau) the set of agent histories consistent with \publicttau.

Importantly, we do not need to encode the mechanisms defining \spott\ in the histories of the agent-extensive form, because the mechanisms are \emph{akin} to a move by nature in the extensive form \extensive(\priort,\actionuptot,\spott). It thus follows that in the agent-extensive form we can define the agent's strategy, \agentstrat, as a measurable function of the private histories.

Finally, we note that the agent evaluates the payoffs of a strategy in  \extensive(\priort,\actionuptot,\spott) using $U(\actionuptot,\cdot,\type)$. We are now ready to define optimal play by the agent in \extensive(\priort,\actionuptot,\spott):

\textbf{Agent-PBE:} Together with the agent strategy \agentstrat, we can also define a system of beliefs $\agentbelief\equiv(\agentbelief_\dateindexb)_{\dateindexb\geq t}^\terminal$, which describes for each period $\dateindexb\geq t$ and for each public history \publicttau, the principal's beliefs over the private histories, $\agentbelief_\dateindexb(\publicttau)\in\Delta\left(\Types\times\Privatettau(\publicttau)\right)$.
%

We say that $\agentassessment$ is an agent-PBE of the agent-extensive form $\extensive(\priort,\actionuptot,\spott)$ if the agent's strategy is sequentially rational (under payoffs $U(\actionuptot,\cdot)$) and the belief system satisfies Bayes' rule where possible. Although the belief system is not needed to test whether the agent's strategy is optimal in the extensive-form game, it is needed to test the optimality of the principal's choice of mechanism.

\autoref{prop:fp-private-history-irrelevance} applies verbatim, allowing us to conclude that for every agent-PBE \agentassessment\ of \extensive(\priort,\actionuptot,\spott), an outcome equivalent \agentassessmentb\ exists, in which the agent's strategy only conditions on her type and the public history. This property is responsible for the recursive nature of the set of PBE-feasible outcomes here and also in the mechanism-selection game.
%
Hereafter, when we say agent-PBE, we mean one that satisfies the above property. 

\textbf{(Continuation) Outcome distributions:} An agent-PBE \agentassessment\ of \extensive(\priort,\actionuptot,\spott) defines a distribution over $\Types\times\allocationsterminal$, $\doutcome^{\spott,\strat}$, that satisfies two conditions. First, the marginal on \Types\ is \priort. Second, the distribution is supported on those tuples $(\type,\action^\terminal)$ such that $\action^\terminal$ coincides with \actionuptot\ through $t-1$. (Online Appendix E.1 contains the formal definition.)

Furthermore, at any history \publicttau, the belief assessment together with the dynamic mechanisms and the agent's strategy, defines a \emph{continuation } outcome, $\doutcome^{\spott,\strat|\publicttau}$, whose marginal on \Types\ coincides with $\agentbelief_\dateindexb(\publicttau)$ and assigns positive probability to those allocations that are consistent with $\actionuptot$ and \publicttau.

\textbf{Principal's sequential rationality:} Fix  a dynamic mechanism given \actionuptot, \spott, and an agent-PBE \agentassessment\ of \extensive(\priort,\actionuptot,\spott). Suppose that the principal considers offering mechanism \spotbt\ instead of $\spot_t$. In order to determine whether the principal wishes to deviate to \spotbt, we need to determine the outcome distributions that can follow \spotbt. We denote this set by $\Deviant_{\eqbmoutcomes_\collection}(\priort,\actionuptot,\spotbt)$ and is defined as follows:
%
\small
\begin{align}\label{eq:deviant}
\Deviant_{\eqbmoutcomes_{\collection}}(\priort,\actionuptot,\spotbt)=\left\{\begin{array}{ll}\doutcomeb\in\Delta(\Types\times\allocationsterminal)&:\doutcomeb=\doutcome^{\spotbtt,\agentstratb}\text{ where} \spotbtt=(\spotbt,\spotbttplus) \text{ and }\\
&\text{(i)}\spotbtt \text{ is a dynamic mechanism given \actionuptot}\\
&\text{(ii)} \agentassessmentb\text{ is an agent-PBE of }\extensive(\priort,\actionuptot,\spotbtt)\\
&\text{(iii)}(\forall (s^\prime,\action^\prime,\corb)\in\saemptyb\times\Cor)\doutcome^{\spotbtt,\agentstratb|s^\prime,\action^\prime,\corb}\in\eqbmoutcomes_{\collection}(\beliefend_{t+1}(s^\prime,\action^\prime,\corb),\actionuptot,\action^\prime)\end{array}\right\}.
\end{align}
\normalsize
In words, an outcome \doutcomeb\ is in $\Deviant_{\eqbmoutcomes_{\collection}}(\cdot,\spotbt)$ if it satisfies two properties. First, $\doutcomeb =\doutcome^{\spotbtt,\agentstratb}$
for a dynamic mechanism \spotbtt\ such that \spotbt\ is the period $t$-mechanism and \agentassessmentb\ is an agent-PBE given \spotbtt. Second, \emph{continuation} outcomes are PBE-feasible. The reason that we are able to require that continuation outcomes are PBE-feasible is that whenever the agent does not condition her strategy on the payoff-irrelevant part of the private history, the following holds: If $\agentassessmentb$ is an agent-PBE of $\extensive(\priort,\actionuptot,\spotbtt)$, then for all $\publicttplus=(s^\prime,\action^\prime,\corb)$, $\agentassessmentb|_{\publicttplus}$ is an agent-PBE of $\extensive(\agentbeliefb_{t+1,\Types}(\publicttplus),(\actionuptot,\action^\prime),(\spotindexb^\prime(\publicttplus,\cdot))_{\dateindexb\geq t+1})$.

While we can use the set $\Deviant_{\eqbmoutcomes_\collection}(\priort,\actionuptot,\cdot)$ to test whether the principal has a deviation from \spott\ at the root of \extensive(\priort,\actionuptot,\spott), \autoref{definition:dm-sequential-rationality} also describes how we test for sequential rationality at later points in the agent-extensive form:
\begin{definition}[Sequential rationality]\label{definition:dm-sequential-rationality} Fix \periodgeqone, $(\priort,\actionuptot)$, a dynamic mechanism \spott\ given \actionuptot, and an agent-PBE \agentassessment\ of \extensive(\priort,\actionuptot,\spott). \spott\ is \emph{sequentially rational} given \agentassessment\ if the following hold:
\begin{enumerate}
\item For all $\spotbt:\messages\mapsto\Delta(\signals\times\allocations)$, a distribution $\doutcomeb\in\Deviant_{\eqbmoutcomes_{\collection}}(\cdot,\spotbt)$ exists such that the principal prefers $\doutcome^{\spott,\agentstrat}$ to \doutcomeb; that is,
$\int_{\Types\times\allocationsterminal}W(\actionuptot,\cdot,\type)d\doutcome^{\spott,\agentstrat}\geq\int_{\Types\times\allocationsterminal}W(\actionuptot,\cdot,\type)d\doutcomeb.$
\item For all $\publicttplus=(s^\prime,\action^\prime,\corb)\in(\saemptyb\times\Cor)$, $\doutcome^{\spott,\agentstrat|\publicttplus}\in\eqbmoutcomes_{\collection}(\beliefend_{t+1}(\publicttplus),\actionuptot,\action^\prime)$.
\end{enumerate}
\end{definition}
The first part of \autoref{definition:dm-sequential-rationality} states that the principal has no deviations in period $t$. The second part says that the principal has no deviations in periods $\dateindexb\geq t+1$: the continuation outcome distribution induced by \spott\ and \agentassessment\ is PBE - feasible in the continuation. 

\autoref{definition:dm-sequential-rationality} resembles the sequential rationality conditions in the mechanism-selection game, except for one important aspect: when we consider the outcomes the principal faces in the event of a deviation, we do not require that the agent's strategy is measurable in any way with respect to the history of mechanisms so far.\footnote{Indeed, we only need the principal's belief \priort\ and the allocations so far \actionuptot\ to describe what is PBE-feasible from period $t$ onward. In a sense, the past mechanisms and output messages are bygones.} By contrast, underlying the definition of the mechanism-selection game is the ability to \emph{measurably} select as a function of \spotbt\ (in fact, as a function of the whole sequence of mechanisms that has been offered through period $t-1$) the outcomes that the principal faces in the event of a deviation.

We are now ready to define the set of PBE feasible outcomes at $(\priort,\actionuptot)$, $\eqbmoutcomes_{\collection}(\priort,\actionuptot)$:

\begin{definition}[PBE-feasible outcomes]\label{definition:pbe-feasible}
Fix \periodgeqone, $(\priort,\actionuptot)\in\Posteriors\times\allocationsuptot$. The distribution $\doutcome\in\Delta(\Types\times\allocationsterminal)$ is PBE-feasible at (\priort,\actionuptot) if a dynamic mechanism \spott\ given \actionuptot\ and an agent-PBE \agentassessment\ of \extensive(\priort,\actionuptot,\spott) exist such that
\begin{enumerate}
\item $\doutcome$ is the outcome distribution induced by \agentassessment\ in $\extensive(\priort,\actionuptot,\spott)$,  $\doutcome^{\spott,\agentstrat}$,
\item \spott\ is sequentially rational given \agentassessment.
\end{enumerate}
$\eqbmoutcomes_{\collection}(\priort,\actionuptot)$ denotes the set of PBE-feasible outcomes at $(\priort,\actionuptot)$.
\end{definition}

By varying \collection, we can define the set of PBE-feasible outcomes when the principal can offer mechanisms whose input and output messages are $(\Types,\Posteriors)$. Like in \autoref{theorem:main-theorem}, our interest is in the canonical-PBE-feasible outcomes, that is, those outcomes that are induced by canonical dynamic mechanisms \cspottt\ (\autoref{definition:canonical-mechanisms}) and canonical-agent PBE of the extensive-form game \extensive(\priort,\actionuptot,\cspottt). In a slight abuse of notation, let $\ceqbmoutcomes(\cdot)$ denote the correspondence of canonical-PBE-feasible outcomes when $\collection=\{(\Types,\Posteriors)\}$.
\begin{theorem}\label{theorem:main-theorem-continuum}
For all $\periodgeqone$ and pairs $(\priort,\actionuptot)\in\Posteriors\times\allocationsuptot$, $\eqbmoutcomes_{\collection}(\priort,\actionuptot)=\ceqbmoutcomes(\priort,\actionuptot)$.
\end{theorem}
The proof of \autoref{theorem:main-theorem-continuum}, which can be found in Online Appendix E.4, follows almost immediately from that in \autoref{theorem:main-theorem}. The only difference is that the proof of \autoref{theorem:main-theorem-continuum} must account for the explicit recursive structure of the sets $\eqbmoutcomes_{\collection}$ and \ceqbmoutcomes. In particular, we are implicitly defining \ceqbmoutcomes\ by requiring that the continuation outcomes are drawn from the (continuation) set \ceqbmoutcomes. However, in accounting for this difference, we illustrate how both sets can be defined in terms of an operator similar to that considered in \cite{abreu1990toward}, which in turn could be used in an application to characterize these sets.


We conclude \autoref{sec:continuum} by illustrating the framework in \autoref{sec:dynamic-mechanisms} within \autoref{ex:sdg}:
\subsection{\autoref{ex:sdg} revisited}\label{sec:sdg}
We now consider \autoref{ex:sdg} under the assumption that the set of types is a continuum. Formally, let $\Types=[\mint,\maxt]\subseteq\mathbb{R}$, for some finite $\mint<\maxt$. In what follows, we use the standard cdf notation \priorf\ and \posteriorf\ to denote the seller's prior and posterior beliefs, instead of \prior\ and $\beliefend_2$ as in \autoref{sec:sdg-program}.

\textbf{Seller's program:}  We first argue that the program \ref{eq:seller-opt} continues to represent the seller's maximum revenue under a PBE-feasible outcome distribution. By \autoref{theorem:main-theorem-continuum}, it continues to be without loss of generality to assume the seller  chooses dynamic canonical mechanisms and to analyze the canonical agent-PBE of the game induced by the canonical dynamic mechanism.
The only difference between the program \ref{eq:seller-opt} and the framework in \autoref{sec:dynamic-mechanisms} is that in \ref{eq:seller-opt} the seller only chooses the \periodone\ mechanism, instead of a dynamic mechanism. However, this distinction is inconsequential: One of the conditions in \autoref{definition:pbe-feasible} is that given the seller's belief \posteriorf, the \periodtwo-outcome distribution is PBE-feasible in \periodtwo. It follows that the \periodtwo-mechanism chosen by the seller must maximize his revenue given his posterior belief about the buyer. Proposition 2 in \cite{skreta2006sequentially} implies the \periodtwo-mechanism is a posted price as a function of \posteriorf, $p_2(\posteriorf)$.

\textbf{Virtual surplus representation:} The assumption of a continuum of types allows us to apply standard mechanism design tools to represent the seller's payoff as a function of the allocation rule and the distribution over posteriors implied by the mechanism. To see this, let $\virtual(\type,\priorf)=\type-(1-\priorf(\type))/\priorpdf(\type)$ denote the buyer's virtual value.

As we show in \autoref{appendix:sdg}, the incentive constraints deliver the envelope representation of the buyer's payoffs, so we can replace the transfers out of the seller's payoff and reduce \ref{eq:seller-opt} to the following program. In period $1$, the seller chooses a distribution over posteriors, $P_{\Posteriors}$, and for each posterior he induces, a probability of trade $q(\posteriorf)$ to solve
\begin{align}\label{eq:sdg-virtual}
\max_{P_{\Posteriors},q}\int_{\Posteriors}\left[q(\posteriorf)\int_{\mint}^{\maxt}\virtual(\type,\priorf)\posteriorf(d\type)+(1-q(\posteriorf))\delta\int_{p_2(\posteriorf)}^{\maxt}\virtual(\type,\priorf)\posteriorf(d\type)\right]P_{\Posteriors}(d\posteriorf)
\end{align}
subject to (i) $P_{\Posteriors}$ must be Bayes' plausible given \priorf\ and (ii) a monotonicity condition, which states that, in expectation, higher types must trade with higher probability  (see \autoref{eq:sdg-monotonicity} in \autoref{appendix:sdg}). \autoref{eq:sdg-virtual} describes the seller's payoff in terms of the distribution over posteriors induced by the mechanism. If at posterior \posteriorf\ the seller sells the good ($q(\posteriorf)=1$), he obtains the expected virtual surplus, where the expectation is calculated using \posteriorf, but the virtual values are calculated using \priorf. This reflects that the probability with which the seller pays rents to a buyer of type \type\ is measured by the probability $\priorf(\type)$ that buyer types below \type\ receive the good. Instead, if at \posteriorf\ the seller does not sell the good ($q(\posteriorf)=0$) he obtains the (discounted) expected virtual surplus of selling the good at price $p_2(\posteriorf)$. Although the posted price in period $2$ is optimal with respect to the \emph{posterior} virtual values  $\virtual(\type,\posteriorf)$, it may not be for the \emph{prior} virtual values. This reflects the conflict between the period $1$ and period $2$ sellers: if they hold different beliefs about the buyer's type, they pay rents with different probabilities, and therefore may prefer different mechanisms.

Although the solution to the \label{page-not-solve-body} problem in \autoref{eq:sdg-virtual} is beyond the scope of this paper,\footnote{\autoref{eq:sdg-virtual} defines an information design problem with a continuum state space where the designer's payoff depends on more than just the posterior mean, and hence is outside the scope of the existing tools in information design (c.f., \citealp{kolotilin2018optimal,dworczak2019simple}).\label{ftn:posterior-mean}} the virtual surplus representation of the seller's problem allows us to expand on what is known about the sale of a durable good under limited commitment. \cite{skreta2006sequentially} shows that, among the mechanisms in \cite{bester2001contracting}, posted-prices are the optimal mechanism for the seller. Instead, when endowed with canonical mechanisms, \autoref{prop:seller-deviation} provides conditions under which the seller will profitably deviate from the optimal posted price mechanism. In other words, the outcome distribution induced by posted prices is not PBE-feasible:
\begin{prop}[Posted prices not PBE-feasible]\label{prop:seller-deviation} Assume $\virtual(\type,\priorf)$ is strictly increasing and $\mint=0$. Then, $\overline{\delta}\in(0,1)$ exists such that for all discount factors $\delta\in(\overline{\delta},1)$, the outcome distribution implemented by posted prices is not PBE-feasible.
\end{prop}
\label{page-ration-body}
The proof of \autoref{prop:seller-deviation} is constructive. Starting from the optimal posted-price mechanism, we show the seller can deviate to an \emph{obfuscated} non-uniform pricing mechanism. This mechanism is characterized by five parameters $(\tilde{\type}_1,\tilde{\type}_2,\gamma,p_\gamma,p_1)$ and works as follows. Pricing is non-uniform because in period $1$ types above $\tilde{\type}_2$ are served with probability $1$ and pay $p_1$, while types in $(\tilde{\type}_1,\tilde{\type}_2)$ are \emph{rationed}, that is, they are served with probability $\gamma$ and pay $p_\gamma$ if they receive the good. The remaining types are not served in period $1$ and pay nothing. It is obfuscated because when $\gamma<1$, the seller observes whether the good is sold in \periodone, but not whether the buyer's type is above or below $\tilde{\type}_1$. Non-uniform pricing has been studied in the durable goods literature under commitment and limited capacity (see \cite{loertscher2019monopoly} and the references therein). Instead, \cite{denicolo1999rationing} show that obfuscation may benefit a durable goods seller if he cannot commit to the sequence of posted prices.\footnote{\cite{denicolo1999rationing} consider a two-period model, where in each period the seller can choose both a price at which to sell the good, and a probability $\gamma$ at which the buyer who wants to buy the good at the posted price, receives the good. Implicit in their analysis is that the seller only observes whether the good is sold but not whether the buyer is willing to buy the good at the posted price.}
%
As we explain next, it is the combination of non-uniform pricing and obfuscation that makes this mechanism more attractive to the seller than the optimal posted-price mechanism.



The proof of \autoref{prop:seller-deviation} shows the optimal posted-price mechanism is, in a sense, costly for the seller when the seller is patient: to avoid offering low prices in \periodtwo, the seller does not trade with buyer types with positive virtual values in \periodone. Instead, by carefully designing the interval $(\tilde{\type}_1,\tilde{\type}_2)$, the obfuscated non-uniform pricing mechanism allows the seller to serve these buyer types with positive probability in \periodone, without necessarily lowering the price in \periodtwo. The proof constructs a deviation to such a mechanism that guarantees that the unique best response for the buyer is to participate and truthfully report her type. It follows that in this setting the set of continuation outcomes described in \autoref{eq:deviant} is not large enough to deter the seller from this deviation.

We conclude \autoref{sec:sdg} by discussing the reason that, unlike \cite{skreta2006sequentially}, other mechanisms within our class can outperform posted prices in a two-period-setting:
%
\begin{rem}[Comparison with \citealp{skreta2006sequentially}]\label{rem:vs}
To understand why in our two-period example the seller can do better than by using the optimal posted price mechanism,
 it is instructive to compare the incentive constraints in \ref{eq:seller-opt} with those implied by mechanisms where the seller observes the buyer's choice of input message as in, for instance, \cite{laffont1988dynamics,bester2001contracting,skreta2006sequentially}.
%
%
%
Although not expressed in the language of type reports or beliefs, the incentive constraints in \cite{skreta2006sequentially} require that for each \posteriorf\ in the support of $\beta(\cdot|\type)$, the buyer prefers the tuple $(q(\posteriorf),\transfer(\posteriorf),u^*(\type,\posteriorf))$ to any other tuple $(q(\posteriorfb),\transfer(\posteriorfb),u^*(\type,\posteriorfb))$ in the mechanism. In particular, the buyer must be indifferent between any two tuples that she chooses with positive probability. Contrast this with the incentive constraints in \ref{eq:seller-opt}, where the buyer is not necessarily
indifferent between the tuples $(q(\posteriorf),\transfer(\posteriorf),u^*(\type,\posteriorf))$ in the support of $\beta(\cdot|\type)$, although in expectation, the lottery she faces over such tuples under truthtelling must be better than the one she faces by lying. Indeed, in the obfuscated non-uniform pricing mechanism, the seller exploits the weaker incentive constraints in \ref{eq:seller-opt}: Buyer types in $(\newcut_1,\newcut_2)$ are not indifferent between receiving the good in period $1$ at the rationing price and receiving the good in period $2$.

However, for longer horizons, the comparison of the seller's payoffs in the two models is not obvious because the larger set of canonical mechanisms also implies the seller has a larger set of deviations in our model than in the model in \cite{skreta2006sequentially}.
\end{rem}

The previous discussion highlights that under limited commitment, the principal may benefit from employing mechanisms where the output message (and hence, the allocation) does \emph{not} reveal the input message that the agent submitted into the mechanism. Contrast this to the standard revelation principle for the case of commitment when the principal faces a privately informed agent (adverse selection): as we explained in the introduction, it follows from the result in \cite{myerson1982optimal} that it is without loss of generality in that case to consider mechanisms whereby the principal learns the input message from observing the realization of the output message. Instead, \cite{myerson1982optimal} shows that adding ``noise'' to the communication may be essential when the principal also faces an agent whose actions are not contractible (moral hazard). Indeed, pooling in the same output message different types of the privately informed agent to incentivize the agent whose action is not contractible to follow the recommendation may be beneficial. Mechanism design with limited commitment is closer to the hybrid model of adverse selection and moral hazard in \cite{myerson1982optimal} than it is to the model of pure adverse selection. Indeed, note that in a given period, the principal faces, in a sense, \emph{two} agents whose incentives he needs to manage: the privately informed agent (adverse selection) and his future self, whose choice of mechanism is not contractible (moral hazard). That is, today's principal needs to elicit the agent's information while simultaneously ensuring his future behavior is sequentially rational. 
In the same way that output messages are key in the presence of moral hazard in \cite{myerson1982optimal}, they feature prominently in our framework. 

Following \cite{myerson1982optimal}, \label{page-self-lang} it would have been natural to consider mechanisms where output messages encode a recommended sequence of mechanisms from tomorrow onward. However, the language of recommendations is self-referential because the set of output messages would refer to continuation mechanisms, which are themselves defined by a set of output messages. Instead, the language of posterior beliefs avoids this potential infinite-regress problem, allowing us to identify a canonical set of output messages for mechanism design with limited commitment.

\section{Conclusions and Further Directions}\label{sec:conclusions}
This paper provides a revelation principle for dynamic mechanism-selection games in which the designer can only commit to short-term mechanisms. In doing so, it opens the door to the study of the implications of limited commitment in fundamental problems in economics and political economy, such as optimal taxation, redistribution, and the design of social insurance (\citealp{sleet2008politically,farhi2012non,golosov2021social}), or environmental regulation (\citealp{hiriart2011weak}), which, due to the difficulties with the revelation principle, have only been studied within simple informational environments, such as fully non-persistent private information. Since our model allows for non-separable payoffs and non-transferable utility, our results can be used in a broad range of applications.
%

A cornerstone of the analysis is the idea that a mechanism should encode not only the rules that determine the allocation, but also the information the designer obtains from the interaction with the agent. We expect that the idea that the mechanism's output messages should encode at the very least the principal's information about the agent is more far-reaching and carries to
other forms of limited commitment, such as renegotiation with long-term mechanisms, where versions of the revelation principle have proved equally elusive.
However, other aspects of our analysis may not carry immediately to the analysis of renegotiation: indeed, \autoref{prop:fp-private-history-irrelevance} uncovered that PBE outcomes have a recursive structure. In games with time-separable payoffs, \autoref{prop:fp-private-history-irrelevance} implies PBE payoffs can be characterized by relying on self-generating techniques as in \cite{abreu1990toward} and \cite{athey2008collusion}, reducing the analysis of a complex dynamic game essentially to a series of static problems (see \citealp{doval2020optimal} for an application). 

%

%

At the same time, by highlighting the canonical role of beliefs as the signals employed by the mechanism, our work opens up a new avenue for research in information design.
 Indeed, as the analysis in \autoref{sec:sdg} and our companion work, \cite{doval2021purchase}, highlights, developing  information design tools for continuum type spaces when the designer's payoff does not depend only on the posterior mean would contribute to our understanding of classical problems in mechanism design.

\bibliographystyle{ecta}
\bibliography{laura-added,limited_commitment_revision,math}
\appendix
\section{Collected definitions and notation}\label{appendix:definitions}
\autoref{appendix:definitions} introduces the necessary notation to define the payoffs from an assessment and, hence, the definition of Perfect Bayesian equilibrium. It also collects notation that is used in the proofs. Throughout this section and also for most of \autoref{appendix:propositions}, we assume that \Types\ is finite and the mechanisms used by the principal have finite support and defer the proof of \autoref{theorem:main-theorem} for the general case to the Online Appendix. 

\textbf{Shorthand notation:} To simplify notation, we do not explicitly include the agent's decision to participate in the mechanism in the histories of the game. Instead, we follow the convention that if the agent does not participate, the input message is \nmssg, the output message is \nsignal, and the allocation is \outsideoption. With this convention in mind, we let $\messages_{i\emptyset}=\messages_i\cup\{\nmssg\}$, $\msaempty\equiv\left(\messages_i\times\signals_j\times\allocations\right)\cup\{(\nmssg,\nsignal,\outsideoption)\}$ and $\saempty\equiv\left(\signals_j\times\allocations\right)\cup\{(\nsignal,\outsideoption)\}$, denote the private and public outcomes in a given period when the principal uses a mechanism with labels $(\messages_i,\signals_j)$. 

Also, given a mechanism $\mechanismt$, let $\zpt$ denote the tuple $\mechanismt,s_t,\actiont$, which summarizes the period-$t$ outcomes from offering \mechanismt, where $(s_t,\actiont)\in\outputt\allocations_\emptyset$. Note that any public history at the beginning of period $t$ can be written as $\publict=(\publictminus,z_{(s_{t-1},\action_{t-1})}(\mechanismtminus),\cor_t)$, with the convention that when $\periodone$, $h^1=\{\cor_1\}$ for some $\cor_1\in[0,1]$. Finally, given an assessment, \assessment, it is useful to collapse the distribution on $\inputt\outputt\allocations_\emptyset$, defined by
\small
\begin{align}\label{eq:kernel}
(1-\participate_t(\type,\hagt,\mechanismt))\mathbbm{1}[(m_t,\signalt,\actiont)=(\nmssg,\nsignal,\outsideoption)]+\participate_t(\type,\hagt,\mechanismt)\comm_t(\type,\hagt,\mechanismt)(m_t)\varphit(\signalt,\actiont|m_t)
\end{align}
\normalsize
and we denote it by $\kernel_t^{\strat}(m_t,s_t,a_t|\tripletfp)$. 

\textbf{Perfect Bayesian equilibrium:} To define Perfect Bayesian equilibrium, we need to define the principal and the agent's payoff from a given assessment. To do so, fix an assessment, \assessment. The prior \prior\ and the strategy profile $\plainstrat=(\stratp,\strat)$ induce a probability distribution over the terminal histories $\Types\times\Hagterminal$, $P^\plainstrat$, via the Ionescu-Tulcea theorem (\citealp{pollard2002user}).\footnote{Online Appendix D provides a formal definition of this distribution.} Moreover, fixing $t$ and $(\type,\hagt)$, the measure $P^{\plainstrat|\type,\hagt}$ corresponds to the measure induced by drawing with probability 1 $(\type,\hagt)$ and then using the continuation strategy profile to determine the distribution over the continuation histories. Fix a public history \publict. 
The principal's payoff at $\publict$ is given
by\small
\begin{align*}
W(\plainstrat,\beliefend|\publict)&=\sum_{\type,\hagt\in\Hagt(\publict)}\beliefend_t(\type,\hagt|\publict)\mathbb{E}_{\stratpt(\publict)}\left[\mathbb{E}^{P^{\plainstrat|(\type,\hagt,\mechanismt)}}\left[W(\actionuptot,\cdot,\type)\right]\right]
\equiv\mathbb{E}_{\stratpt(\publict)}\left[W(\plainstrat,\beliefend|\publict,\mechanismt)\right],
\end{align*}\normalsize
where $\actionuptot=(\action_1,\dots,\actiontminus)$ is the allocation through the beginning of period $t$ that is consistent with \publict\ (with the convention that $\action^0=\{\emptyset\}$). 
Note that the principal's payoff from offering \mechanismt\ at \publict, $W(\plainstrat, \beliefend|\publict,\mechanismt)$, depends on the belief system \beliefend\ only through $\beliefend_t(\cdot|\publict)$. 

For any mechanism \mechanismt, $W(\plainstrat,\beliefend|\publict,\mechanismt)$ equals
\begin{align*}
\sum_{\type,\hagt\in\Hagt(\publict)}\beliefend_t(\type,\hagt|\publict)\sum_{(m_t,\signalt,\actiont)\in\inputt\outputt\allocations_\emptyset}\kernel_t^{\strat}(m_t,s_t,\actiont|\type,\hagt,\mechanismt)\mathbb{E}^{P^{\plainstrat|(\type,\hagt,\mechanismt,m_t,s_t,a_t)}}[W(\actionuptot,\actiont,\cdot,\type)].
\end{align*}
%
The principal's beliefs at \publict, together with mechanism \mechanismt\ and the agent's strategy, define a distribution over the continuation public histories as follows:
\begin{align}\label{eq:public-kernel}
\nu_{t+1}^{\beliefend,\strat}(\publict,\zpt|\publict,\mechanismt)&=\sum_{\type,\hagt\in\Hagt(\publict),m_t\in\inputt_\nmssg}\beliefend_t(\type,\hagt|\publict)\kernel_t^{\strat}(m_t,\signalt,\actiont|\tripletfp).
\end{align}
With this notation in hand, we can express the principal's payoff $W(\plainstrat,\beliefend|\publict,\mechanismt)$ as:
\begin{align}\label{eq:fp-principal-payoff}
W(\plainstrat,\beliefend|\publict,\mechanismt)&=\sum_{(s_t,\actiont)\in\outputt\allocations_\emptyset}\nu_{t+1}^{\beliefend,\strat}(\publict,\zpt|\publict,\mechanismt)W(\plainstrat,\beliefend|\publict,\zpt).
\end{align}
Similarly, the agent's payoff at private history $(\type,\hagt)$, when the principal offers mechanism \mechanismt, is given by:
\small
\begin{align}\label{eq:fp-agent-payoff}
U(\plainstrat|\type,\hagt,\mechanismt)&=
\sum_{(m_t,\signalt,\actiont)\in\inputt\outputt\allocations_\emptyset}\kernel_t^{\strat}(m_t,s_t,\actiont|\tripletfp)\mathbb{E}^{P^{\plainstrat|(\type,\hagt\mechanismt,m_t,\signalt,\actiont)}}\left[U(\actionuptot,\actiont,\cdot,\type)\right].
\end{align}\normalsize
With this, we can formally define Perfect Bayesian equilibrium. 
\begin{definition}\label{definition:sequential-rationality} An assessment \assessment\ is \emph{sequentially rational} if for all $\periodgeqone$ and public histories $\publict$, the following hold:
\begin{enumerate}
\item If $\space\space\mechanismt$ is in the support of \stratpt(\publict), then $W(\plainstrat,\beliefend|\publict,\mechanismt)\geq W(\plainstrat,\beliefend|\publict,\mechanismbt)$ for all $\mechanismbt\in\mechanismscollection$,
\item For all $(\type,\hagt)\in\Types\times\Hagt(\publict)$, and \space\mechanismt\ in \mechanismscollection, $U(\plainstrat|\type,\hagt,\mechanismt)\geq U(\stratp,\stratb|\type,\hagt,\mechanismt)$ for all $\stratb$.
\end{enumerate}
\end{definition}
\begin{definition}\label{definition:brwp} An assessment \assessment\ \emph{satisfies Bayes' rule where possible} if
for all public histories \publict\ and mechanisms \mechanismt\ the following holds:
\begin{align}
&\beliefend_{t+1}(\type,\hagt,m_t,\zpt,\cor_{t+1}|\publict,\zpt,\cor_{t+1})\nu_{t+1}^{\beliefend,\strat}(\publict,\zpt|\publict,\mechanismt)\label{eq:bayes-p}\\
&=\beliefend_t(\type,\hagt|\publict)\kernel_t^{\strat}(m_t,\signalt,\actiont|\type,\hagt,\mechanismt)\nonumber.
\end{align}
\normalsize
\end{definition}
\begin{definition} An assessment $(\stratp,\strat,\beliefend)$ is a \emph{Perfect Bayesian equilibrium} if it is sequentially rational and satisfies Bayes' rule where possible.
\end{definition}

\textbf{Prunning:} Given a mechanism \mechanismt, let 
$(\outputt\times\allocations)_+=\{(s_t,\actiont):(\exists m\in\inputt)\varphit(s_t,\actiont|m)>0\}.$
The set $\outputt\times\allocations\setminus(\outputt\times\allocations)_+$ has zero probability \emph{regardless} of the agent's strategy. Hence, if we remove from the tree those paths that are consistent with mechanism \mechanismt\ and $(s,\action)\notin(\outputt\times\allocations)_+$, this does not change the set of equilibrium outcomes. Hereafter, these histories are removed from the tree.

\textbf{Principal pure strategies: } Lemma D.1 in Online Appendix D.2 shows that without loss of generality, we can focus on PBE assessments of \gamecollection\ in which the principal plays a pure strategy, so in what follows, we focus on PBE assessments that satisfy Lemma D.1.
\section{Proof of \autoref{theorem:main-theorem}}\label{appendix:propositions}
The proof of \autoref{theorem:main-theorem} follows from the proof of Propositions \ref{prop:fp-private-history-irrelevance}-\ref{prop:fp-bijection} below.

\begin{prop}\label{prop:fp-private-history-irrelevance}
For every PBE assessment \assessment\ of \gamecollection, an outcome-equivalent PBE assessment \assessmentb\ exists such that the agent's strategy only depends on her type and the public history.
\end{prop}
We relegate the proof of \autoref{prop:fp-private-history-irrelevance} to Online Appendix D.1.1. In what follows, we focus on PBE of the mechanism-selection game that satisfy the properties of \autoref{prop:fp-private-history-irrelevance} and abuse notation in the following two ways: First, we write the agent's strategy as a function of her private type and the public history alone, with the understanding that $\stratt(\type,\hagt)=\stratt(\type,\publict)$ whenever $\hagt\in\Hagt(\publict)$. Similarly, we write the belief system at history \publict\ as inducing distributions over $\Types$ and not over $\Types\times\Hagt(\publict)$.

\begin{prop}\label{prop:fp-bijection}
Fix \collection\ and let \assessment\ be a PBE assessment of \gamecollection\ that satisfies  \autoref{prop:fp-private-history-irrelevance}. Then, an outcome-equivalent canonical PBE assessment \assessmentb\ of \gamecollection\ exists.
\end{prop}
\begin{proof}[Proof of \autoref{prop:fp-bijection}]
Let \assessment\ be as in the statement of \autoref{prop:fp-bijection}. Let $\publict$ be a public history and let \mechanismt\ denote the mechanism that the principal offers at $\publict$ under \stratpt.  Let $\Types^+$ denote the support of the principal's beliefs at \publict, $\beliefend_t(\publict)$.

For types in $\Types^+$, use \autoref{eq:kernel} to define an auxiliary mapping $\varphi^\prime:\Types^+\mapsto\Delta\left(\outputt\allocations_\emptyset\right)$, as follows:
\begin{align}\label{eq:auxiliary}
&\varphi^\prime(\signalt,\actiont|\type)=\sum_{m\in\inputt_\nmssg}\kernel_t^{\strat}(m_t,\signalt,\actiont|\type,\publict,\mechanismt).
\end{align}
That is, $\varphi^\prime$ corresponds to the \emph{direct} version of \varphit\ for $\type\in\Types^+$; we use it in what follows to construct an alternative mechanism for the principal, \mechanismbt, that uses message sets $(\Types,\Posteriors)$.

Omitting the dependence on (\strat, \posterior), recall that $\nu_{t+1}(\publict,\zpt|\publict,\mechanismt)$ denotes the probability of history $(\publict,\zpt)$ under the equilibrium strategy when the principal offers \mechanismt\ at \publict\ (\autoref{eq:public-kernel}). \autoref{eq:public-kernel} implies we can write $\beliefbeg_{t+1}$ using \spotb\ as follows:
\begin{align*}
\beliefbeg_{t+1}(\publict,\zpt|\publict,\mechanismt)&=\sum_{\type\in\Types^+}\beliefend_t(\type|\publict)\varphi^\prime(s_t,\actiont|\type).
\end{align*}
In what follows, to simplify notation we omit the dependence of $\nu_{t+1}(\cdot|\publict,\mechanismt)$ on \mechanismt. 

The first step is to show that the distribution over continuation histories $\nu_{t+1}$ can be seen as inducing a distribution over posterior beliefs, allocations, and realizations of a public randomization device. To see this, for $\posterior\in\Posteriors$, let $\Belief(\posterior)$ denote the set
\begin{align*}
\Belief(\posterior)=\left\{(s_t,\actiont)\in\outputt\allocations_\emptyset:\posterior_{t+1}(\cdot|\publict,\zpt)=\posterior\right\},
\end{align*}
and let $\Belief_{\actiont}(\posterior)$ denote the projection of \Belief(\posterior) onto $\{\actiont\}$. In what follows, for any subset $\Belief\subseteq\outputt\allocations_\emptyset$, we abuse notation and write $\nu_{t+1}(\Belief|\publict)$ instead of $\sum_{(s_t,\actiont)\in\Belief}\nu_{t+1}(\publict,\zpt|\publict)$. 

Let $\Posteriors^+$ denote the smallest subset of \Posteriors\ such that $\nu_{t+1}(B(\Posteriors^+)|\publict)=1$. That is, $\Posteriors^+$ is the set of principal posterior beliefs that are pinned down via Bayes' rule. We can write the principal's payoff at history \publict\ when he offers \mechanismt\ as follows:
\begin{align}\label{eq:principal-payoff}
&\sum_{(s_t,a_t)}\nu_{t+1}(\publict,\zpt|\publict)W(\sigma,\posterior_{t+1}(\cdot|\publict,\zpt)|\publict,\zpt)\\
&=\sum_{\posterior\in\Posteriors^+}\nu_{t+1}(\Belief(\posterior)|\publict)\sum_{\actiont\in\allocations}\underbrace{\frac{\nu_{t+1}(\Belief_{\actiont}(\posterior)|\publict)}{\nu_{t+1}(\Belief(\posterior)|\publict)}}_{\text{allocation rule}}\sum_{\signalt\in\Belief_{\actiont}(\posterior)}\underbrace{\frac{\nu_{t+1}(\publict,\zpt|\publict)}{\nu_{t+1}(\Belief_{\actiont}(\posterior)|\publict)}}_{\text{public randomization}}W(\sigma,\posterior|\publict,\zpt).\nonumber
\end{align}
The above equation shows two ways in which we can think of the distribution over continuation histories starting from \publict. The first is standard: we draw history (\publict,\zpt) using the distribution induced by the equilibrium strategy, $\beliefbeg_{t+1}(\cdot|\publict)$. The second is the one that delivers the canonical mechanisms: we first draw a belief \posterior\ using the distribution over continuation equilibrium beliefs induced by $\beliefbeg_{t+1}(\cdot|\publict)$ and then we draw an allocation \actiont, conditional on the continuation equilibrium belief coinciding with \posterior. The principal's posterior belief \posterior\ and the allocation \actiont\ may still not be enough to pin down the continuation history, so we draw the output message $s_t$ conditional on $s_t$ being consistent with \actiont\ and \posterior. 
%

The second step is to show that, conditional on the induced posterior belief \posterior, (i) the probability that the allocation is \actiont\ is independent of \type, and (ii) the probability that the output message is $s_t\in\Belief_{\actiont}(\posterior)$ is independent of \type. To see this, note that for any belief $\posterior\in\Posteriors^+$, for any $(s_t,\actiont)\in\Belief(\posterior)$ and for any $\type$ such that $\posterior(\type)>0$, we have\footnote{Note that if $\posterior\in\Posteriors^+$ and $\type$ is such that $\posterior(\type)>0$, then $\type\in\Types^+$.}\small
\begin{align}\label{eq:bayes-rule}
\posterior(\type)&=\frac{\posterior_t(\type|\publict)\varphi^\prime(s_t,\actiont|\type)}{\nu_{t+1}(\publict,\zpt|\publict)}=\frac{\posterior_t(\type|\publict)\sum_{s_t^\prime\in\Belief_{\actiont}(\posterior)}\varphi^\prime(s_t^\prime,\actiont|\type)}{\nu_{t+1}(\Belief_{\actiont}(\posterior)|\publict)}
=\frac{\posterior_t(\type|\publict)\sum_{(s_t^\prime,\actiont^\prime)\in\Belief(\posterior)}\varphi^\prime(s_t^\prime,\actiont^\prime|\type)}{\nu_{t+1}(\Belief(\posterior)|\publict)}.
\end{align}\normalsize
That is, the principal updates to \posterior\ either when (i) he observes $(s_t,\actiont)\in\Belief(\posterior)$, (ii) he learns that \actiont\ is the realized allocation, that is, he learns that $s_t\in\Belief_{\actiont}(\beliefend)$, or (iii) he learns that the output message and the allocation belong to \Belief(\posterior). Thus, for all \type\ in the support of \posterior, we have
\begin{align}\label{eq:constant-lr}
\frac{\nu_{t+1}(\Belief_{\actiont}(\posterior)|\publict)}{\nu_{t+1}(\Belief(\posterior)|\publict)}&=\frac{\sum_{s_t^\prime\in\Belief_{\actiont}(\posterior)}\spotb(s_t^\prime,\actiont|\type)}{\sum_{(s_t^\prime,\actiont^\prime)\in\Belief(\posterior)}\spotb(s_t^\prime,\actiont^\prime|\type)}
\\
\frac{\nu_{t+1}(\publict,\zpt|\publict)}{\nu_{t+1}(\Belief_{\actiont}(\posterior)|\publict)}&=\frac{\spotb(s_t,\actiont|\type)}{\sum_{s_t^\prime\in\Belief_{\actiont}(\posterior)}\spotb(s_t^\prime,\actiont|\type)},\nonumber
\end{align}
where each of the equalities follows from applying \autoref{eq:bayes-rule}. \autoref{eq:constant-lr} shows (i) the probability that the allocation is \actiont\ conditional on the induced belief being \posterior\ is independent of \type, and (ii) the probability that the output message is $s_t\in B_{\actiont}(\posterior)$ conditional on the allocation being \actiont\ and the induced belief \posterior\ is independent of \type. It follows that for all $\posterior\in\Posteriors^+$, $(s_t,\actiont)\in\Belief(\posterior)$ and for all \type\ in the support of \posterior, we can \emph{split} the auxiliary mapping as follows:
\begin{align*}
\varphi^\prime(s_t,\actiont|\type)&=\left(\sum_{(s_t^\prime,\actiont^\prime)\in\Belief(\posterior)}\varphi^\prime(s_t^\prime,\actiont^\prime|\type)\right)\frac{\nu_{t+1}(\Belief_{\actiont}(\posterior)|\publict)}{\nu_{t+1}(\Belief(\posterior)|\publict)}\frac{\nu_{t+1}(\publict,\zpt|\publict)}{\nu_{t+1}(\Belief_{\actiont}(\posterior)|\publict)}
\\&=\frac{\posterior(\type)}{\posterior_t(\type|\publict)}\nu_{t+1}(\Belief(\posterior)|\publict)\frac{\nu_{t+1}(\Belief_{\actiont}(\posterior)|\publict)}{\nu_{t+1}(\Belief(\posterior)|\publict)}\frac{\nu_{t+1}(\publict,\zpt|\publict)}{\nu_{t+1}(\Belief_{\actiont}(\posterior)|\publict)},
\end{align*}
where the last equality follows from the last equality in \autoref{eq:bayes-rule}.

Thus, the agent's payoff at history \publict, when the principal offers mechanism \mechanismt\ and her type is $\type\in\Types^+$, can be written as follows:
\footnotesize
\begin{align}\label{eq:agent-payoff}
&\sum_{(s_t,\actiont)\in\outputt\allocations_\emptyset}\varphi^\prime(s_t,\actiont|\type)\mathbb{E}^{P^{\sigma|(\type,\publict,\zpt)}}\left[U(\actionuptot,\actiont,\cdot,\type)\right]=\\
&\sum_{\posterior\in\Posteriors}\frac{\posterior(\type)}{\posterior_t(\type|\publict)}\nu_{t+1}(\Belief(\posterior)|\publict)\sum_{\actiont\in\allocations}\frac{\nu_{t+1}(B_{\actiont}(\posterior)|\publict)}{\nu_{t+1}(\Belief(\posterior)|\publict)}\sum_{\signalt\in\Belief_{\actiont}(\posterior)}\frac{\nu_{t+1}(\publict,\zpt|\publict)}{\nu_{t+1}(\Belief_{\actiont}(\posterior)|\publict)}\mathbb{E}^{P^{\sigma|(\type,\publict,\zpt)}}[U(\actionuptot,\actiont,\cdot,\type)].\nonumber
\end{align}
\normalsize
The difference between the principal and the agent's payoff in Equations \ref{eq:principal-payoff} and \ref{eq:agent-payoff} is that the agent cares only about the distribution over $(\posterior,s_t,\actiont)$ \emph{conditional} on \type, whereas the principal's payoff is expressed in terms of the unconditional distribution. For this reason the agent's payoff features the term $\posterior(\type)/\posterior_t(\type|\publict)$.

We now define the canonical mechanism $\mechanismt^C=(\Types,\Posteriors,\varphi^{\mechanismt^C})$: First, for $\type\in\Types^+$, 
\begin{align*}
&\varphi^{\mechanismt^C}(\posterior,\actiont|\type)=\underbrace{\frac{\posterior(\type)}{\posterior_{t}(\type|\publict)}\nu_{t+1}(\Belief(\mu)|\publict)}_{\beta^{\mechanismt^C}(\posterior|\type)}\underbrace{\frac{\nu_{t+1}(\Belief_{\actiont}(\posterior)|\publict)}{\nu_{t+1}(\Belief(\posterior)|\publict)}}_{\alpha^{\mechanismt^C}(\actiont|\posterior)},
\end{align*}
where the decomposition in terms of $\beta^{\mechanismt^C},\alpha^{\mechanismt^C}$ is well-defined because of the independence properties highlighted after \autoref{eq:constant-lr}. Second, if $\type\notin\Types^+$, let $\type^*(\type)$ denote the solution to
\small
\begin{align}\label{eq:value-participate}
\max_{\typec\in\Types^+}\sum_{(\posterior,\actiont)\in\Posteriors\times\allocations}\varphi^{\mechanismt^C}(\posterior,\actiont|\typec)\sum_{s_t\in\Belief_{\actiont}(\posterior)}\frac{\beliefbeg_{t+1}(\publict,\zpt|\publict)}{\beliefbeg_{t+1}(\Belief_{\actiont}(\posterior)|\publict)}\mathbb{E}^{P^{\plainstrat|(\type,\publict,\zpt)}}\left[U(\actionuptot,\actiont,\cdot,\type)\right]
\end{align}\normalsize
and let $\varphi^{\mechanismt^C}(\cdot|\type)=\varphi^{\mechanismt^C}(\cdot|\type^*(\type))$. Change the principal's strategy at \publict\ so that he offers $\mechanismt^C$ instead of \mechanismt. Change the agent's strategy so that, conditional on participating, the agent truthfully reports her type, $\commb_t(\type,\publict,\mechanismt^C)=\delta_\type$.

For $\posterior\in\Posteriors^+$ and allocation \actiont, enumerate the output messages in $\Belief_{\actiont}(\posterior)$ as $s_t^1,\dots,s_t^K$. (We omit the dependence of $K$ on \posterior\ and \actiont\ to simplify notation.) Define the sequence $\{\omega_k\}_{k=0}^K$ such that $\omega_0=0,\omega_K=1$ and for $k=1,\dots,K-1$,
\begin{align*}
\omega_k-\omega_{k-1}=\frac{\nu_{t+1}(\publict,z_{(s_t^k,\actiont)}(\mechanismt)|\publict)}{\nu_{t+1}(\Belief_{\actiont}(\posterior)|\publict)}.
\end{align*}
Modify the continuation strategies as follows: for $k=1,\dots,K$ and $\cor\in[\omega_{k-1},\omega_k]$, let $\plainstrat|_{(\publict,z_{(\posterior,\actiont)}(\mechanismt^C),\cor)}$ coincide with $\plainstrat|_{(\publict,z_{(s_t^k,\actiont)}(\mechanismt),\frac{\omega-\omega_{k-1}}{\omega_k-\omega_{k-1}})}$. Note these strategies imply the principal and the agent's payoffs remain the same as in the original equilibrium whenever $\type\in\Types^+$. Furthermore, modify the continuation strategies so that $\plainstrat|_{(\publict,z_{(\emptyset,\outsideoption)}(\mechanismt^C))}=\plainstrat|_{(\publict,z_{(\emptyset,\outsideoption)}(\mechanismt))}$.

For \type\ in $\Types^+$, set $\pi_t^\prime(\type,\publict,\mechanismt^C)=1$. For types not in $\Types^+$, use \autoref{eq:value-participate} to compute $\participateb_t(\type,\publict,\mechanismt^C)$ accordingly. Conditional on participating, the agent can guarantee at most the payoff from imitating the strategy followed by \typed\ for some $\typed\in\Types^+$. This strategy was already feasible in the original PBE, so the agent has no new deviations. It follows that the new assessment is a PBE of the auxiliary game.
\end{proof}

\subsection{Proof of \autoref{prop:finite-support}}\label{appendix:finite}
%

Fix \collection\ and let \assessment\ denote a canonical PBE of \gamecollection, like the one constructed in \autoref{prop:fp-bijection}. Fix a history \publict\ and let \mechanismt\ denote the mechanism the principal offers at \publict\ under \stratpt. Let $\Types^+$ denote the support of $\beliefend_t(\publict)$. To simplify notation, we use the shorthand notation  $(w^*,(u_\type^*)_{\type\in\Types^+},(u_\type^{\participate})_{\type\notin\Types^+})$ to denote the principal and the agent's payoff vector at \publict\ when the agent participates in \mechanismt, $(W(\plainstrat,\beliefend|\publict,\mechanismt),U(\plainstrat|\type,\publict,\mechanismt,1))$. \autoref{theorem:main-theorem} implies that for the principal and for $\type\in\Types^+$, $w^*$ and $u_\type^*$ are their equilibrium payoffs at \publict.
%

Similar steps to those in the proof of \autoref{prop:fp-bijection} in Online Appendix D.1 imply a distribution $\tau\in\Delta(\Posteriors)$ exists such that for all measurable subsets \measurablem\ of \Posteriors\ and for all $\type\in\Types$, 
\begin{align*}
\betat(\measurablem|\type)\posterior_t(\type|\publict)=\int_{\measurablem}\posteriorb(\type)\tau(d\posteriorb).
\end{align*}
Therefore, we can write the principal and the agent's payoffs as follows:
\begin{align*}
w^*&=\int_{\Posteriors}\sum_{\type\in\Types}\posteriorb(\type)\left[\int_{\allocations}\mathbb{E}^{P^{\plainstrat|(\type,\publict,\mechanismt,\posteriorb,\actiont)}}\left[W(\actionuptot,\actiont,\cdot,\type)\right]\alphat(d\actiont|\posteriorb)\right]\tau(d\posteriorb)\equiv\int_{\Posteriors} w(\posteriorb)\tau(d\posteriorb),\\
u_\type^*&=\int_{\Posteriors}\int_{\allocations}\mathbb{E}^{P^{\plainstrat|(\type,\publict,\mechanismt,\posteriorb,\actiont)}}\left[U(\actionuptot,\actiont,\cdot,\type)\right]\alphat(d\actiont|\posteriorb)\frac{\posteriorb(\type)}{\posterior_t(\type|\publict)}\tau(d\posteriorb)\equiv\int_{\Posteriors}u(\posteriorb,\type)\tau(d\posteriorb),\\
u_{\type}^{\participate}&=\int_{\Posteriors}\int_{\allocations}\mathbb{E}^{P^{\plainstrat|(\type,\publict,\mechanismt,\posteriorb,\actiont)}}\left[U(\actionuptot,\actiont,\cdot,\type)\right]\alphat(d\actiont|\posteriorb)\frac{\posteriorb(\type^*(\type))}{\posterior_t(\type^*(\type)|\publict)}\tau(d\posteriorb)\equiv\int_{\Posteriors}u(\posteriorb,\type)\tau(d\posteriorb),
\end{align*}
where $\type^*(\type)\in\Types^+$ is as in the proof of \autoref{prop:fp-bijection} (recall \autoref{eq:value-participate}).
Finally, note that the payoff of the agent of type \type\ from reporting $\typed\in\Types^+$ at history \publict\ when the mechanism is \mechanismt, and then following her equilibrium strategy is given by:
\begin{align*}
&\int_{\Posteriors}\int_{\allocations}\mathbb{E}^{P^{\plainstrat|(\type,\publict,\mechanismt,\posteriorb,\actiont)}}\left[U(\actionuptot,\actiont,\cdot,\type)\right]\alphat(d\actiont|\posteriorb)\betat(d\posteriorb|\typed)\\
&=\int_{\Posteriors}\int_{\allocations}\mathbb{E}^{P^{\plainstrat|(\type,\publict,\mechanismt,\posteriorb,\actiont)}}\left[U(\actionuptot,\actiont,\cdot,\type)\right]\alphat(d\actiont|\posteriorb)\frac{\posteriorb(\typed)}{\posterior_t(\typed|\publict)}\tau(d\posteriorb)\equiv\int_{\Posteriors}\deviation_\type(\typed,\posteriorb)\tau(d\posteriorb)=\deviation_{\type,\typed}^*.
\end{align*}
The notation $\deviation_{\type}$ denotes that these are the payoffs that \type\ obtains from deviating to \typed. The payoff $\deviation_{\type,\typed}^*$ can be similarly defined for $\typed\notin\Types^+$ using $\type^*(\typed)$ as in the definition of $u_\type^\participate$.

 Let $S=\text{supp}\tau$. Define
$ C=\{(\posteriorb,w(\posteriorb),(u(\type,\posteriorb))_{\type\in\Types},(\deviation_\type(\typed,\posteriorb))_{\type,\typed\in\Types}):\posteriorb\in S\}.$
Because \Types\ is finite, \cite{rubin1958note} implies that $(\posterior(\publict),w^*,(u_\type^*)_{\type\in\Types^+},(u_\type^\participate)_{\type\notin\Types^+},(\deviation_{\type,\typed}^*)_{\type,\typed\in\Types})\in\text{co}C$. Letting $N=|\Types|$, Caratheodory's theorem implies $M\leq N(N+1)+1$ exists such that $(\posterior(\publict),w^*,(u_\type^*)_{\type\in\Types^+},(u_\type^\participate)_{\type\notin\Types^+},(\deviation_{\type,\typed}^*)_{\type,\typed\in\Types})$ can be written as the convex combination of $M$ elements of $C$. That is, $(\lambda_i,\posteriorb_i)_{i=1}^M$ exist such that  $\lambda_i\geq 0,\sum_{i=1}^M\lambda_i=1$, and \small
 \begin{align*}
 \posterior_t(\publict)&=\sum_{i=1}^M\lambda_i\posteriorb_i
 &w^*&=\sum_{i=1}^M\lambda_iw(\posteriorb_i)
& u_\type^*&=\sum_{i=1}^M\lambda_iu(\type,\posteriorb_i)
& u_\type^\participate&=\sum_{i=1}^M\lambda_iu(\type,\posteriorb_i)
 &\deviation_{\type,\typed}^*&=\sum_{i=1}^M\lambda_i\deviation_\type(\typed,\posteriorb_i).
 \end{align*}\normalsize
Consider the following canonical mechanism \mechanismbt: For types in $\Types^+$, let $\betabt(\{\posteriorb_i\}|\type)=\frac{\posteriorb_i(\type)}{\posterior_t(\type|\publict)}\lambda_i$, and otherwise, $\betabt(\cdot|\type)=0$. For types not in $\Types^+$, let $\betabt(\cdot|\type)=\betabt(\cdot|\type^*(\type))$. Furthermore, $\alphabt(\cdot|\posteriorb_i)=\alphat(\cdot|\posteriorb_i)$. Modify the PBE assessment so that the principal offers $\mechanismbt$ at \publict. Furthermore, modify the continuation strategies so that  $\assessment|_{(\publict,\mechanismbt,z_{(\posteriorb_i,\actiont)}(\mechanismbt))}=\assessment|_{(\publict,\mechanismt,z_{(\posteriorb_i,\actiont)}(\mechanismt))}$ and $\assessment|_{(\publict,\mechanismbt,z_{(\nsignal,\outsideoption)}(\mechanismbt))}=\assessment|_{(\publict,\mechanismt,z_{(\nsignal,\outsideoption)}(\mechanismt))}$. Clearly,  mechanism \mechanismbt\ delivers the same payoff to the principal and the agent conditional on the agent participating and truthfully reporting her type. We now show it is optimal for the agent to truthfully report her type. Suppose the agent of type \type\ reports instead that her type is $\typed\in\Types^+$ (the case $\typed\notin\Types^+$ is similar). In this case, she obtains:\small
\begin{align*}
&\sum_{i=1}^M\lambda_i\frac{\posteriorb_i(\typed)}{\posterior_t(\typed|\publict)}\int_{\allocations}\mathbb{E}^{P^{\plainstrat|\type,\publict,z_{(\posteriorb_i,\actiont)}(\mechanismbt)}}\left[U(\actionuptot,\actiont,\cdot,\type)\right]\alphabt(d\actiont|\posteriorb_i)=\sum_{i=1}^M\lambda_i\deviation_\type(\typed,\posteriorb_i)=\deviation_{\type,\typed}^*
\end{align*}\normalsize
Since $u_\type^*\geq \deviation_{\type,\typed}^*$, then truthtelling is still optimal. Thus, because the payoffs from participating and not participating are the same as in the original assessment for all types,
$\participate(\type,\publict,\mechanismbt)=\participate_t(\type,\publict,\mechanismt)$ is a best response.

\section{Proofs of \autoref{sec:sdg}}\label{appendix:sdg}
\textbf{Envelope theorem:} We establish the envelope theorem and the virtual surplus representation of the seller's payoff. The buyer's payoff  in the mechanism when her type is \type\ is given by
\begin{align}\label{eq:buyer-payoff}
U(\type)=\int_{\Posteriors}\left(\type q(\posteriorf)-\transfer(\posteriorf)+(1-q(\posteriorf))\delta u^*(\type,\posteriorf)\right)\beta(d\posteriorf|\type),
\end{align}
where $u^*(\type,\posteriorf)=\max\{0,\type -p_2(\posteriorf)\}$. 
\autoref{eq:deviation-type} defines the buyer's payoff when her type is \type\ and she deviates by first reporting \typed\ and then following \typed's strategy in \periodtwo:
\small
\begin{align}\label{eq:deviation-type}
V(\typed,\type)=\int_{\Posteriors}\left(\type q(\posteriorf)-\transfer(\posteriorf)+(1-q(\posteriorf))\delta \left[u^*(\typed,\posteriorf)+(\type-\typed)q_2^*(\typed,\posteriorf)\right]\right)\beta(d\posteriorf|\typed),
\end{align}
\normalsize
where $q_2^*(\typed,\posteriorf)=\mathbbm{1}[\typed\geq p_2(\posteriorf)]$.
The optimality of truthtelling implies
$U(\type)=\max\left\{V(\typed,\type):\typed\in\Types\right\}$.
We now establish that the family $\{V(\typed,\cdot):\typed\in\Types\}$ is equi-Lipschitz continuous. To see this, let $\type,\typec$ such that $\typec\neq\type$ and consider
\begin{align*}
&|V(\typed,\type)-V(\typed,\typec)|=\left|\int_{\Posteriors}\left[(\type-\typec)q(\posteriorf)+(1-q(\posteriorf))\delta (\type-\typec)q_2^*(\typed,\posteriorf)\right]\beta(d\posteriorf|\typed)\right|\\
&\leq|\type-\typec|\left|\int_{\Posteriors}\left[q(\posteriorf)+(1-q(\posteriorf))\delta q_2^*(\typed,\posteriorf)\right]\beta(d\posteriorf|\typed)\right|\leq|\type-\typec|(1+\delta).
\end{align*}
Moreover, for \type\ in the interior of \Types\ we have that
\begin{align*}
\frac{\partial}{\partial\type}V(\typed,\type)=\lim_{h\rightarrow 0}\frac{V(\typed,\type+h)-V(\typed,\type)}{h}=\int_{\Posteriors}\left[q(\posteriorf)+(1-q(\posteriorf))\delta q_2^*(\typed,\posteriorf)\right]\beta(d\posteriorf|\typed).
\end{align*}
$U(\type)$ is then Lipschitz continuous because it is the maximum over a family of equi-Lipschitz continuous functions.
Thus, it is differentiable almost everywhere. Theorem 1 in \cite{milgrom2002envelope} implies that at any point of differentiability of $U(\cdot)$, 
\begin{align*}
U^\prime(\type)=\frac{\partial}{\partial\typec}V(\type,\typec)|_{\typec=\type}=\int_{\Posteriors}\left[q(\posteriorf)+(1-q(\posteriorf))\delta q_2^*(\type,\posteriorf)\right]\beta(d\posteriorf|\type).
\end{align*}
It follows that\small
\begin{align}\label{eq:transfers}
\int_{\Types}\left[\int_{\Posteriors}\transfer(\posteriorf)\beta(d\posteriorf|\type)\right]&\priorf(d\type)=\int_{\Types}\int_{\Posteriors}\left[\type q(\posteriorf)+(1-q(\posteriorf))\delta u^*(\type,\posteriorf)\right]\beta(d\posteriorf|\type)\priorf(d\type)\nonumber\\
&-\int_{\Types}\int_{\mint}^{\type}\left[\int_{\Posteriors}\left(q(\posteriorf)+(1-q(\posteriorf))\delta q_2^*(u,\posteriorf)\right)\beta(d\posteriorf|u)\right]du\priorf(d\type),
\end{align}\normalsize
where we are already replacing that at the optimum, transfers will be chosen so that $U(\mint)=0$. Recall from \ref{eq:seller-opt} that the seller's revenue is given by
$\mathbb{E}_{\priorf}\left[\int_{\Posteriors}\left(\transfer(\posteriorf)+(1-q(\posteriorf))\delta p_2(\posteriorf)q_2^*(\type,\posteriorf)\right)\beta(d\posteriorf|\type)\right]$.
 Replacing the transfers (\autoref{eq:transfers}) and integrating by parts, we obtain:
\begin{align}\label{eq:step2}
\int_{\Types}\int_{\Posteriors}\left[q(\posteriorf)\virtual(\type,\priorf)+(1-q(\posteriorf))\delta q_2^*(\type,\posteriorf)\virtual(\type,\priorf)\right]\beta(d\posteriorf|\type)\priorf(d\type).
\end{align}
Denote by $P$ the distribution on $\Types\times\Posteriors$ defined as
$P(\measurablet\times\measurablem)=\int_{\measurablet}\beta(\measurablem|\type)\priorf(d\type)$, 
for all measurable subsets $\measurablet,\measurablem$ of $\Types$ and $\Posteriors$. Let $P_{\Posteriors}$ denote its marginal on \Posteriors, \citet[Appendix F, Theorem 6]{pollard2002user} implies \autoref{eq:step2} equals
\begin{align*}
\int_{\Posteriors}\left[q(\posteriorf)\int_{\Types}\virtual(\type,\priorf)\posteriorf(d\type)+(1-q(\posteriorf))\delta\int_{p_2(\posteriorf)}^{\maxt}\virtual(\type,\priorf)\posteriorf(d\type)\right]P_{\Posteriors}(d\posteriorf),
\end{align*}
which is the expression in \autoref{eq:sdg-virtual}.

\textbf{Monotonicity:} Together with the envelope representation of the buyer's payoffs, the mechanism must also satisfy the following monotonicity constraint for all $\type\geq\typed$:
\begin{align}\label{eq:sdg-monotonicity}
\int_{\Posteriors}\left[(\type-\typed)q(\posteriorf)+\delta(1-q(\posteriorf))\int_{\typed}^{\type}q_2^*(u,\posteriorf)du\right]\left(\beta(d\posteriorf|\type)-\beta(d\posteriorf|\typed)\right)\geq 0.
\end{align}
\normalsize
\begin{proof}[Proof of \autoref{prop:seller-deviation}]
We first establish properties that the optimal posted-price mechanism satisfies. Let $\posteriorf^{\cut}$ denote the distribution of \type\ conditional on $\type\leq\cut$. Since $\virtual(\type,\priorf)$ is strictly increasing, it follows that $\virtual(\type,\posteriorf^{\cut})$ is also strictly increasing. Furthermore, since $\mint=0$, $\virtual(\mint,\posteriorf^{\cut})<0<\virtual(\cut,\posteriorf^{\cut})$ and there exists a unique $p_2(\cut)$ such that $\virtual(p_2(\cut),\posteriorf^{\cut})=0$. It follows that $p_2(\cut)=\arg\max_{p_2} p_2(1-\posteriorf^{\cut}(p_2))$. The theorem of the maximum implies that $p_2(\cut)$ is continuous in \cut. Thus, the optimal posted-price mechanism solves
\begin{align}\label{eq:posted-prices-value}
R(\delta)=\max_{\cut}\int_{\cut}^{\maxt}\virtual(\type,\priorf)\priorf(d\type)+\delta\int_{p_2(\cut)}^{\cut}\virtual(\type,\priorf)\priorf(d\type).
\end{align}
The theorem of the maximum implies that a solution exists. Note that $R(0)=R(1)=\int_{p_M}^{\maxt}\virtual(\type,\priorf)\priorf(d\type)$, where $p_M$ is the monopoly price, that is, $\virtual(p_M,\priorf)=0$. For each $\delta\in(0,1)$, let $\cut_\delta^*$ denote a solution to the problem in \autoref{eq:posted-prices-value} and define
 \[V(\delta^\prime,\delta)=\int_{\cut_{\delta^\prime}^*}^{\maxt}\virtual(\type,\priorf)\priorf(d\type)+\delta\int_{p_2(\cut_{\delta^\prime}^*)}^{\cut_{\delta^\prime}^*}\virtual(\type,\priorf)\priorf(d\type)\equiv\int_{\cut_{\delta^\prime}^*}^{\maxt}\virtual(\type,\priorf)\priorf(d\type)+\delta C(\delta^\prime).\]
Note that $R(\delta)=\sup_{\delta^\prime\in[0,1]}V(\delta^\prime,\delta)$. Furthermore, $R(\delta)\geq V(\delta^\prime,\delta)$ and $R(\delta^\prime)\geq V(\delta,\delta^\prime)$ imply that if $\delta>\delta^\prime$, then $C(\delta^\prime)\leq C(\delta)$. Note that $C(0)<0<C(1)=R(1)$ since the unique solution at $\delta=0$ is to set $\cut_0^*=p_M$. Steps similar to those leading to the envelope representation of the buyer's payoffs imply that (i) $\{V(\delta^\prime,\cdot):\delta^\prime\in[0,1]\}$ is equi-Lipschitz continuous, (ii) for all $\delta^\prime\in[0,1]$, $V(\delta^\prime,\cdot)$ is differentiable on $(0,1)$ with derivative equal to $C(\cdot)$, and (iii) $R$ is Lipschitz continuous and at any point of differentiability, $R^\prime(\delta)=C(\delta)$. 

We now show that $\overline{\delta}\in(0,1)$ exists so that $C(\delta)>0$ for $\delta\in(\overline{\delta},1]$. Toward a contradiction, suppose that $C(\delta)\leq0$ for all $\delta\in(0,1)$. Then, $R$ is decreasing on $(0,1)$ and because $R(0)=R(1)$ it follows that $C(\cdot)\equiv0$ almost everywhere. Otherwise, a set $D\subseteq(0,1)$ of non-zero Lebesgue measure exists such that $C(u)<0$ for $u\in D$. Then,
\begin{align*}
R(1)=R(0)+\int_0^1C(u)du=R(0)+\int_{[0,1]\setminus D}C(u)du+\int_DC(u)du\leq R(0)+\int_DC(u)du<R(0),
\end{align*}
where the first inequality follows from $C(u)\leq 0$ on $(0,1)$ and the second by the definition of $D$.  It follows that $C\equiv 0$ on $(0,1)$ and hence $R(\cdot)$ is constant and equal to the commitment payoff for all $\delta\in(0,1)$, a contradiction. Thus,  a subset of $(0,1)$ exists on which $C$ is strictly positive and because $C$ is increasing,  $\overline{\delta}\in(0,1)$ exists so that $C(\delta)>0$ for $\delta\in(\overline{\delta},1]$.

Second, we use the above properties to construct a deviation for the seller. Define $\overline{\delta}\in(0,1)$ as above and let $\delta>\overline{\delta}$. We show  $\epsilon,\eta,\gamma>0$ exist such that the seller can deviate to the following mechanism. Let $\priceopt\equiv p_2(\cutopt)$ and define three posterior beliefs:
\begin{align*}
\posteriorsale^1(\type)&=\frac{\priorf(\type)-\priorf(\cut_\delta^*)}{1-\priorf(\cut_\delta^*)}\mathbbm{1}[\type\geq\cut_\delta^*], \quad
\posteriorsale^\epsilon(\type)=\frac{\priorf(\type)-\priorf(\priceopt-\eta)}{\priorf(\cut_\delta^*)-\priorf(\priceopt-\eta)}\mathbbm{1}\left[\type\in[\priceopt-\eta,\cut_\delta^*]\right]\\
\posteriordelay(\type)&=\left\{\begin{array}{ll}\frac{\priorf(\type)}{\priorf(\priceopt-\eta)+(1-\epsilon)(\priorf(\cut_\delta^*)-\priorf(\priceopt-\eta))}&\text{ if }\type<\priceopt-\eta
\\
\frac{\priorf(\priceopt-\eta)+(1-\epsilon)(\priorf(\type)-\priorf(\priceopt-\eta))}{\priorf(\priceopt-\eta)+(1-\epsilon)(\priorf(\cut_\delta^*)-\priorf(\priceopt-\eta))}&\text{ if }\type\in[\priceopt-\eta,\cut_\delta^*]\end{array}\right.,
\end{align*}
and let $q(\posteriorsale^1)=q(\posteriorsale^\epsilon)=1$ and $q(\posteriordelay)=0$. Furthermore, let $\beta(\{\posteriorsale^1\}|\type)=\mathbbm{1}[\type>\cutopt]$, and let
\begin{align}
\beta(\{\posteriordelay\}|\type)=\left\{\begin{array}{ll}1&\text{ if } \type<\priceopt-\eta\\
1-\epsilon&\text{ if } \type\in[\priceopt-\eta,\cut_\delta^*]\\
0&\text{otherwise}\end{array}\right.,
&\beta(\{\posteriorsale^\epsilon\}|\type)=\left\{\begin{array}{ll}0&\text{ if } \type<\priceopt-\eta\\
\epsilon&\text{ if } \type\in[\priceopt-\eta,\cut_\delta^*]\\
0&\text{otherwise}\end{array}\right..\nonumber
\end{align}
Finally, for $\gamma>0$, define payments as follows:
\begin{align*}
\transfer(\posteriordelay)=-\gamma\epsilon,\transfer(\posteriorsale^\epsilon)=-\gamma\epsilon+\priceopt-\eta,\transfer(\posteriorsale^1)=-\gamma\epsilon+\epsilon(\priceopt-\eta)+(1-\epsilon)((1-\delta)\cut_\delta^*+\delta\priceopt).
\end{align*}
That is, the seller now serves with probability $\epsilon$ all types in $[\priceopt-\eta,\cut_\delta^*]$, and leaves the probability of trade for the remaining types unchanged; furthermore, he pays the buyer $\gamma\epsilon$, regardless of her type. The value of $\eta$  is chosen so that $\priceopt$ remains optimal in period $2$. 

In what follows, we first argue that for any agent-extensive form that is feasible given the above mechanism in period $1$ and in any agent-assessment of that extensive form, the buyer must accept the above mechanism with probability 1. That, conditional on participating, truthtelling is optimal is immediate. We then show that given $\epsilon$ we can always find $\eta$ so that \priceopt\ remains optimal in period $2$. Finally, we show that for $\epsilon,\gamma$ small, the seller prefers the mechanism defined above to the optimal posted-price mechanism.

\textbf{The buyer participates with probability $1$:} Note first that buyer types in $[\mint,\mint+\gamma\epsilon)$ must accept the mechanism: the best price they can obtain in period $2$ is \mint, whereas they obtain at least $\gamma\epsilon$ by participating in the mechanism offered by the seller. Since $\gamma\epsilon>\type-\mint$ for all such types, they must accept the mechanism. It then follows that, conditional on rejecting the mechanism, the price is at least $\gamma\epsilon+\mint$ in period $2$, so that it must be the case that types in $[\mint+\gamma\epsilon, \mint+2\gamma\epsilon)$ must also accept the mechanism. Proceeding inductively, it follows that the buyer accepts the mechanism with probability $1$. Thus, Bayes' rule does not pin down  beliefs conditional on non-participation, so that we can specify them arbitrarily: In particular, we can specify them so that the seller assigns probability $1$ to \maxt.

\textbf{\priceopt\ remains optimal in period $2$:} The virtual values under $\posteriordelay$ are given by
\begin{align*}
\virtual(\type,\posteriordelay)=\left\{\begin{array}{ll}\type-\frac{(1-\epsilon)\priorf(\cutopt)+\epsilon\priorf(\priceopt-\eta)-\priorf(\type)}{\priorpdf(\type)}&\text{ if }\type\leq\priceopt-\eta\\
\virtual(\type,\posteriorf^{\cutopt})&\text{ if } \type>\priceopt-\eta\end{array}\right.,
\end{align*}
so that the virtual values under \posteriordelay\ (i) are piecewise increasing in \type\ on $[\mint,\cutopt]$, (ii) are (weakly) negative on $(\priceopt-\eta,\priceopt]$, and (iii) they jump down at $\type=\priceopt-\eta$, whenever $\epsilon>0$. Let $H(\eta,\epsilon)$ denote the limit from below of $\virtual(\type,\posteriordelay)$ as $\type\rightarrow \priceopt-\eta$. That is, 
\begin{align*}\label{eq:limit-vv}
\objective(\eta,\epsilon)\equiv \priceopt-\eta-\frac{(1-\epsilon)(\priorf(\cut_\delta^*)-\priorf(\priceopt-\eta))}{\priorpdf(\priceopt-\eta)}.
\end{align*}
Using the strict monotonicity of the virtual values and that $\mint=0$, it is immediate to show that for each $\epsilon>0$, $\eta(\epsilon)$ exists such that $\objective(\eta(\epsilon),\epsilon)=0$. Thus, for $\eta=\eta(\epsilon)$, $\virtual(\type,\posteriordelay)$ is (weakly) negative for types \type\ below $\priceopt-\eta$, and hence \priceopt\ remains optimal. In what follows, it is important to note $\eta(\cdot)$ is increasing in $\epsilon$ and the definition of \priceopt\ implies $\eta(0)=0$.

\textbf{The seller has a deviation:} We now verify that for small enough $\epsilon,\gamma$ the seller prefers the above mechanism to the optimal posted price mechanism. Indeed, the difference in payoffs between the new mechanism and the optimal posted-price mechanism is as follows:
\begin{align*}
\epsilon\left[(1-\delta)\int_{\priceopt}^{\cut_\delta^*}\virtual(\type,\priorf)\priorf(d\type)+\int_{\priceopt-\eta(\epsilon)}^{\priceopt}\virtual(\type,\priorf)\priorf(d\type)-\gamma\right].
\end{align*}
By assumption, the first term in the square brackets is positive and independent of $\epsilon,\eta,\gamma$, whereas for small $\epsilon$, the second term, although negative, is vanishing, and $\gamma$ is also small. Thus, for $\delta>\overline{\delta}$, small enough $\epsilon,\gamma$ exist such that the seller has a deviation.
 \end{proof}
 
\end{document}